\documentclass[final,5p,times,twocolumn]{elsarticle}

\usepackage[T1]{fontenc}     
\usepackage[utf8]{inputenc}  
\usepackage{lmodern}         
\usepackage{amsmath}
\usepackage{mathtools}
\usepackage{xcolor}
\usepackage{enumitem}
\usepackage{comment}
\usepackage{graphicx}
\usepackage{multirow}
\usepackage{array}
\usepackage{caption}
\usepackage{fontawesome5}
\usepackage{hyperref}
\usepackage{wrapfig}
\usepackage{longtable}
\usepackage{times}
\usepackage{latexsym}
\usepackage{geometry}
\usepackage{array}
\usepackage{latexsym}
\usepackage{color, colortbl}
\usepackage{booktabs}
\usepackage{longtable}  
\usepackage{caption} 
\usepackage{float}

\usepackage{amssymb}

\begin{document}
\begin{frontmatter}




\title
{Toxicity in Online Platforms and AI Systems: A Survey of Needs, Challenges, Mitigations, and Future Directions}


\author[1]{Smita Khapre} 
\author[1]{Melkamu Abay Mersha}
\author[1]{Hassan Shakil}
\author[2]{Jonali Baruah}
\author[1]{Jugal Kalita}

\affiliation[1]{organization={College of Engineering and Applied Science, University of Colorado Colorado Springs},
            postcode={80918},
            state={CO},
            country={USA}}


\affiliation[2]{organization={Tarleton State University, A Member of The Texas A \& M University System},
             city={Stephenville},
             postcode={76401}, 
             state={TX},
             country={USA}}

\begin{abstract}
The evolution of digital communication systems and the designs of online platforms have inadvertently facilitated the subconscious propagation of toxic behavior. Giving rise to reactive responses to toxic behavior.
Toxicity in online content and Artificial Intelligence  Systems has become a serious challenge to individual and collective well-being around the world. It is more detrimental to society than we realize. Toxicity, expressed in language, image, and video, can be interpreted in various ways depending on the context of usage. Therefore, a comprehensive taxonomy is crucial to detect and mitigate toxicity in online content, Artificial Intelligence systems, and/or Large Language Models in a proactive manner. A comprehensive understanding of toxicity is likely to facilitate the design of practical solutions for toxicity detection and mitigation. The classification in published literature has focused on only a limited number of aspects of this very complex issue, with a pattern of reactive strategies in response to toxicity. This survey attempts to generate a comprehensive taxonomy of toxicity from various perspectives. It presents a holistic approach to explain the toxicity by understanding the context and environment that society is facing in the Artificial Intelligence era. 
This survey summarizes the toxicity-related datasets and research on toxicity detection and mitigation for Large Language Models, social media platforms, and other online platforms, detailing their attributes in textual mode, focused on the English language. Finally, we suggest the research gaps in toxicity mitigation based on datasets, mitigation strategies, Large Language Models, adaptability, explainability, and evaluation.
\end{abstract}


\end{frontmatter}



\section{Introduction}
The rise of online platforms, social media, and Artificial Intelligence (AI) has revolutionized how people communicate, share information, and express opinions \cite{gao2020public, villate2024systematic}. While these technologies provide significant benefits, they are also the source of toxicity, like hate speech, cyberbullying, misinformation, and harassment \cite{lin2024combating}. The pervasive nature of toxicity in online platforms and AI systems impacts individuals, society, and businesses, and has regulatory implications \cite{kiritchenko2021confronting}. 

Many synonymous terms are used in industry and academia to refer to toxicity in digital media. A partial list includes words and phrases such as hate, hate speech, offensive language, flaming, incivility, risk, malicious, harmful, hateful sexual, illegal and abusive content, information hazards, discrimination, bias, fake news, misinformation, human-computer-interaction harms, and cyberbullying \cite{kirk2022hatemoji, wang2024Donotanswer}. Toxicity is further categorized into implicit and explicit, based on target groups, political and social bias, illegal, violent, aggressive, and more \cite{pachinger2023toward}. These are most prevalent on social media and other online platforms. Toxicity and harm have been interpreted as distinct by some research \cite{wang2024Donotanswer}. In this survey, we treat toxicity and harm as synonyms, with similar impacts. Markov et al. \cite{markov2023Moderation} highlighted the challenges in real-world toxicity detection in Generative AI (GenAI). They emphasized the importance of a well-motivated taxonomy of toxic content being fundamental to their detection. 

The research for detecting the toxicity in online platforms and AI Systems has picked up pace due to its ubiquitous and adverse psychological impacts on society. Misinformation and fake news have a dramatic effect on public opinion and are likely to move society toward extremism \cite{lazer2018science, gjerazi2023impact}. Although individuals of all generations are negatively impacted, the influence on adolescents' faculties seems to be most pronounced. \cite{lemaire2025development}. 

\begin{figure}[h!]
    \centering
    \includegraphics[width=0.9\linewidth]{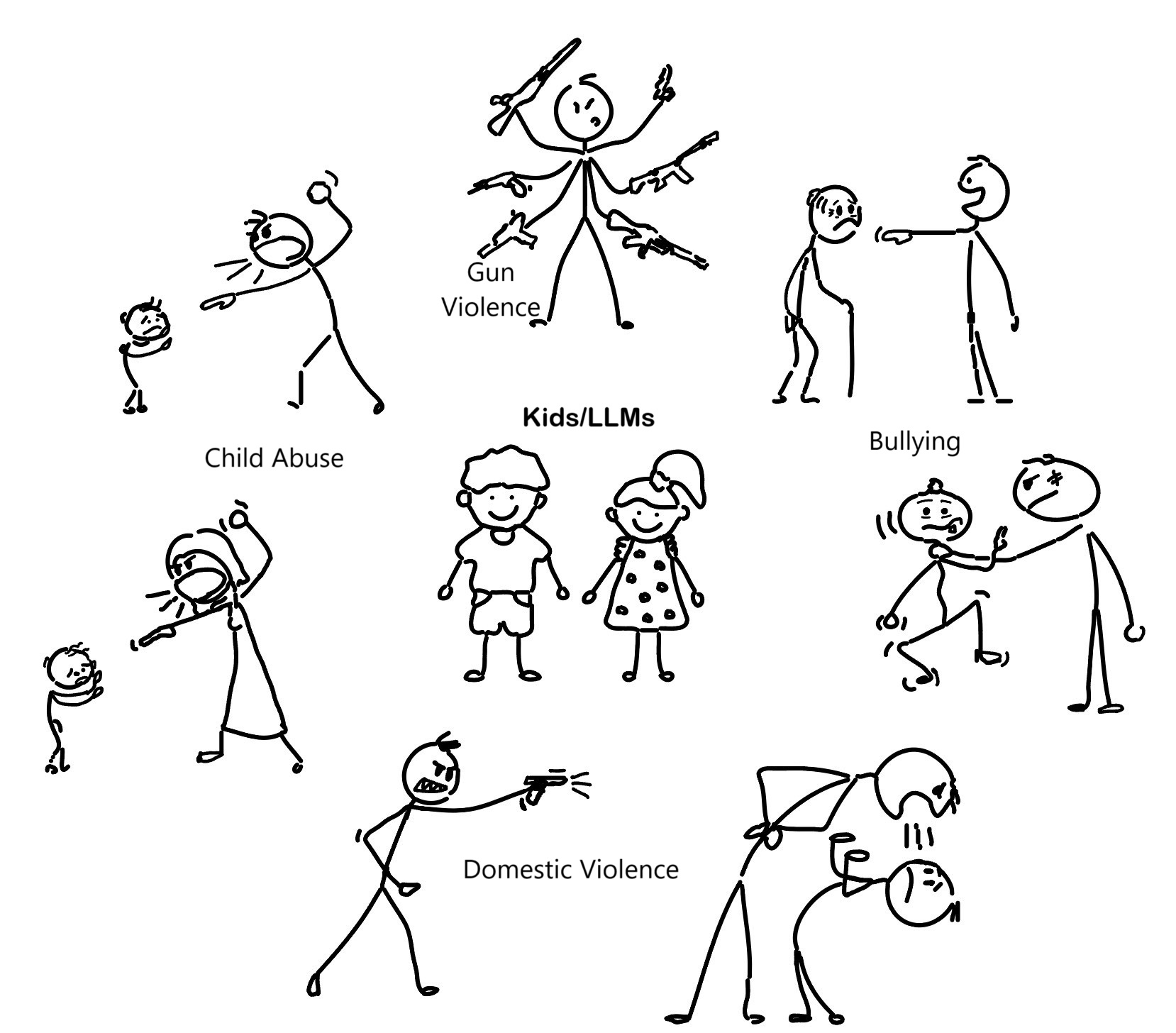}
    \caption{LLMs depicted as Kids, trained on Toxicity Ignorant Datasets, unable to distinguish toxic language from normal language. Just as kids grow up in their toxic environments unaware that it is toxic, when grow up as adults start displaying toxic behavior same as their environments}
    \label{fig:env}
\end{figure}

Platform economy has taken over communication, and most platforms either are beset with toxicity issues or are prone to harbor toxicity in various ways \cite{villate2024systematic}. Large Language Models (LLMs) can be leveraged to detect toxicity in online platforms and detoxify them \cite{gallegos2024bias, mersha2024semantic}. LLMs are trained on very large language corpora. The datasets are usually annotated, and not filtered, containing all types of implicit toxic words, explicit toxic words, and social bias. Thus, LLMs trained on such a dataset also propagate the inherent toxicity of its dataset. An analogical depiction is shown in Figure \ref{fig:env}, where LLMs are depicted as little kids unaware of toxicity, existing in an environment where toxicity abounds. Kids grow up and become an integral part of society with inherent toxic behavior, unaware that it is detrimental and that it should be avoided. LLMs are the underlying technology in GenAI, which generates human-like responses to given prompts.  Toxicity is seen in their responses too. Moreover, using prompt engineering, prompt injection, and jail-breaking attacks, toxicity can be deliberately introduced \cite{wei2024jailbroken, xu2024comprehensive, liu2024arondight}. While there has been tremendous improvement in GenAI, there is a need to implement improved toxicity detection and detoxification \cite{kim2024robust}. 

There is significant research on detection and detoxification of explicit, implicit, and biased toxicities. With current technological advancements, dealing with toxicity requires a careful understanding of the environment and platform. Derave et al. developed a generic module of digital platform (DP) ontology, based on DP attributes, DP attribute values, DP software functionalities, DP types, and DP supported communication \cite{derave2021comparing}. It provides DP design gaps in DPs. The advent of Web 2.0, linguistic toxicity was predominant in communication. Most research dealing with toxicity still focuses only on linguistic expressions, and is usually classified as implicit and explicit toxicity \cite{lewandowska2023integrated}. Their semantics of the words or phrases are used for classification as implicit or explicit. To mitigate the impacts of toxicity comprehensively, we need to understand the environment in which it appears. \textit{Toxicity is a fatal disease that not only requires diagnosis but also cure and prevention.} This metaphor implies that the \textit{Toxicity} is a disease, \textit{diagnosis} is detection, \textit{cure} is detoxification, and \textit{prevention} requires an understanding of the environment and its source and impacts on society. 

While there is tremendous effort in the research community to generate a taxonomy for toxicity in digital communication, with every passing year and the advent of new technology, it is never complete. 
Our survey provides a structured overview of toxicity in online platforms and AI systems. Based on the current state-of-the-art, we examine key areas such as detection methods, mitigation strategies, detoxification techniques, and bias-related concerns. Beyond technical approaches, we also explore the psychological and societal consequences of toxic content, along with the limitations of existing datasets. This work aims to inform future research and support the development of safer, more inclusive digital ecosystems. The main contributions of our survey are as follows.
\begin{itemize}[leftmargin=1.5em]
  \item We present a holistic taxonomy of toxicity, capturing diverse forms of harmful behavior, language, and content, in the context of various platforms, including insights from human psychology. 
  \item We systematically review over 200 studies addressing toxicity detection, mitigation, detoxification, associated biases, and the evolution of digital communication and online platforms.
  \item We evaluate major toxicity datasets identifying key limitations such as bias, annotation inconsistency, and coverage gaps.
  \item We examine the broader psychological and societal impacts of toxic content on individuals and communities.
  \item We identify open challenges and outline future research directions to advance responsible and inclusive online and AI systems.
\end{itemize}

\begin{figure*}[h!]
    \centering
    \includegraphics[width=1 \linewidth]{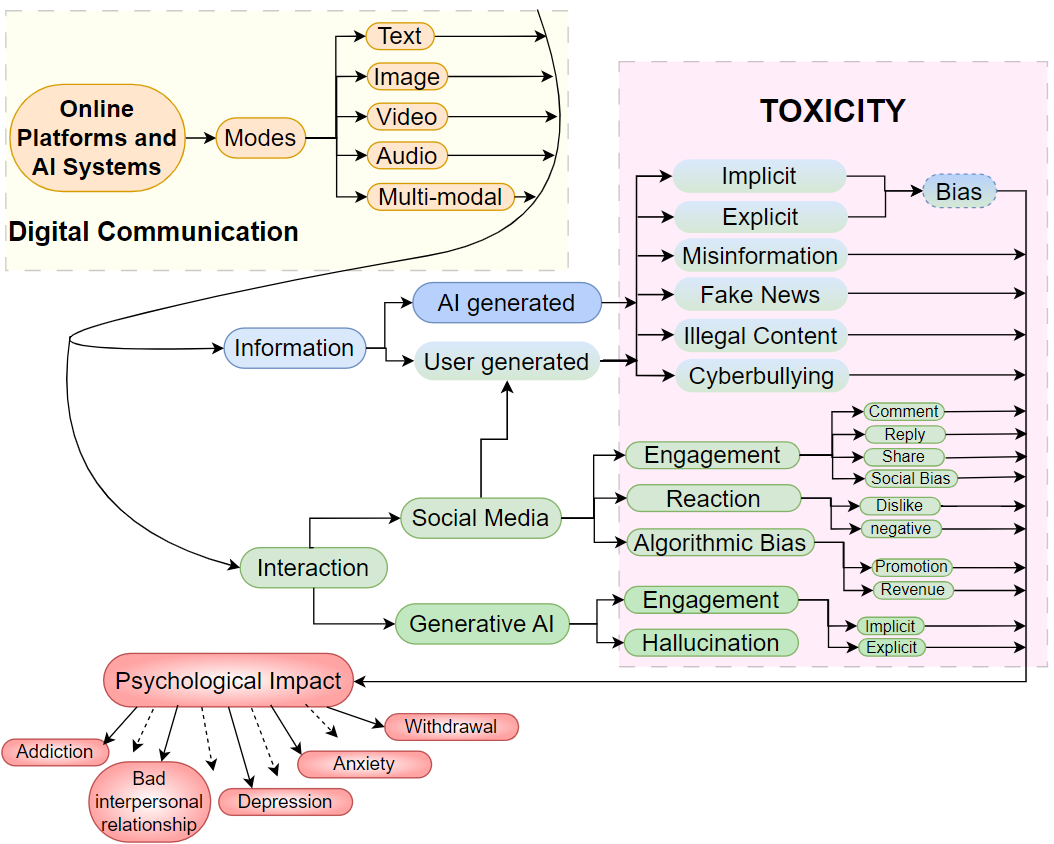}
    \caption{A Taxonomy of Toxic Content based on origination of digital communication in online platforms and AI systems via text, image, audio, video, and/or multi-modal channel. Each mode of communication hosts toxicity divided mainly into information-based and interaction-based systems. Red color signifies the impact of toxicity, green is toxicity in interaction-based platforms, Blue represents other online platforms, and blue-green color gradient in both social media and other online platforms. The pink colored box encompasses the toxicities requiring the detection and mitigation effort}
    \label{fig:T-cls}
\end{figure*}

The remainder of this paper is structured as follows.  In Section \ref{Sec_Bkgd_n_Moti}, we discuss \textit{Background and Motivation} for this survey. It includes the basic terminologies used, why the study of toxicity matters, and outlines the stakeholders of the toxicity ecosystem. In Sections \ref{Sec_Taxonomy} and \ref{Sec_Fine}, we discuss a taxonomy of toxicity we have developed (Figure \ref{fig:T-cls}). In Section \ref{Sec_Psycho}, we examine the impacts of toxicities on human psychology. In Section \ref{Sec_Mitigation}, we collate and discuss the toxicity datasets, followed by detection and detoxification research. In Section \ref{Sec_Open}, we present the open challenges that still need to be addressed and future directions. Finally, in Section \ref{Sec_conclusion}, we draw the conclusion of this survey. 

\section{Background and Motivation} \label{Sec_Bkgd_n_Moti}
Modes of communication have evolved through history. Since 1792, there has been rapid technological progress in communication. Dhingra et al. \cite{dhingra2019historical} navigate the evolution of communication technology from the telegraph in 1792, the telephone in 1876, email (ARPANET) in 1971, the World Wide Web in 1989, and Social Media started in 2003. The emergence of Social Media Platforms (SMP) such as Facebook in 2004, YouTube in 2005, Twitter in 2006, WhatsApp in 2009, Pinterest, and Instagram in 2010 have drastically altered the digital communication landscape. Currently, AI-powered Social Media have taken over channels of communication. Vayadande et al. \cite{Web3} and Shen et al. \cite{shen2024artificialweb1Toweb3} navigate through the evolution and impacts of Web 1.0 (1991-2004), Web 2.0 (2004-2024) to Web 3.0 (2020 onwards). The use of AI improves user experience during enhanced communication, and the use of blockchain technology achieves safety, trust, privacy, and decentralization. Wang et al. \cite{wang2024ToxicityinWeb3} point to the increase in \textit{probability} of toxicity in Web 3.0 content, requiring effective governance policies while we move toward decentralization in Web 3.0. We describe the basic terminologies used before getting into our motivation of this survey.

\subsection{Selected Basic Terminologies}
\begin{itemize}
    \item \textbf{Toxicity} is a term we use to signify a substance or language construct or media element that highly impact an individual in a negative manner. In online context, it is often derogatory, rude, harmful, and disrespectful. Toxic content induces stress, anxiety, suicidal thoughts, depression, and loneliness. These are the least among the impacts on society \cite{akar2025exploring}\cite{maleki2022applying}.
    
    \item \textbf{Implicit Toxicity} is the use of language or media elements that cannot be straightforwardly interpreted as toxic, but which suggest an implied toxic meaning. It introduces toxicity indirectly, which often targets one or more individuals, races, or genders in context. \cite{bkaczkowska2025implicit}. It often involves use of figures-of-speech like metaphor, simile, irony or sarcasm.
    
    \item \textbf{Explicit Toxicity} is the direct and intentional use of toxic language or media elements. It is aggressive and offensive. Bansal et al. describe it as direct and intentional flaming \cite{bansal975classification, lewandowska2023integrated}.
    
    \item \textbf{Fake News} is not authentic news, created intentionally and propagated via diverse users \cite{hu2025overviewFakeNews}. It mimics real news construct. 
    
    \item \textbf{Misinformation} is incorrect information whose authenticity can be verified as false but is generated unintentionally \cite{scheufele2019scienceMisinfo}. It does not mimic real news. Disinformation is similar to misinformation, but is intentionally created false information. A rumor is unverified information that could be true or false. 
    
    \item \textbf{Fallacy} \cite{helwe2024mafalda} provides logical reasoning behind untrue construct or falsehood. It is divided into three main types: ethos (assuming false credibility), logos (false logic), and pathos (appealing to emotions). Fallacies provide support to fake news, misinformation, and disinformation dissemination.
    
    \item \textbf{Illegal Content} is comprised of unauthorized or authorized users selling goods to evade taxes or the sale of smuggled goods, controlled substances, weapons, wildlife, sexually explicit content or trafficking in human, goods, or animals.  Illegal content is unlawful and is regulated by law \cite{dangsawang2024machineSmuggledGoods, mademlis2024invisibleArmsTrafficking, roy2024analysisIllegalWildlife, singh2024ChildAbuseMaterial}.
    
    \item \textbf{Cyberbullying} is bullying or threatening behavior on digital technology platforms \cite{ray2024cyberbullying}. 
    
    \item \textbf{Modes} are the media used to communicate ideas, information, opinions, etc. They include text, audio, and video \cite{chandler2011dictionary}. 
    
    \item \textbf{Engagement} refers to the ability of digital media to keep the user highly interested or engaged.
    
    \item \textbf{Bias} is a false pre-conceived notion based on skewed data or information or expectation assessment or evaluation impacting a target population \cite{delgado2004bias}. 
    
    \item \textbf{Implicit Bias} is an unconscious inclination of one's behavior towards or against others' race, identity, culture, background, color, ability, community, gender, age, sexuality, nationality, status, religion, and so on \cite{banaji2015bias}. 
    
    \item \textbf{Social Bias} is an occurrence of societal discrimination against a target group, person, ideas, or beliefs based on prejudices and stereotypes prevalent in the society \cite{webster2022social}. 
    Sap et al. \cite{sap2020social} developed so-called social bias frames to generate the Social Bias Inference (SBIC) Dataset. 
    
    \item \textbf{Algorithmic Bias} occurs when an algorithm outputs favors or ignores particular target groups or individuals for the wrong reasons, causing imbalanced impacts \cite{kordzadeh2022algorithmic}. It can be divided in two categories. One is when Social Bias, as explained earlier, influences the training dataset and generates the same bias in outputs \cite{johnson2021algorithmic}. The second is an intentional bias built into the platforms to gear the users' engagement and revenue generation. The Netflix Documentary ``The Social Dilemma,'' directed by Jeff Orlowski in 2020, has explained it exceptionally well.  
\end{itemize}

\subsection{Why Does Toxicity Matters?} \label{Sec_Whymatters} 

Understanding the implications of online toxicity is crucial to address and develop effective mitigation strategies. The implications include psychological harm, social division, proliferation of misinformation, ethical challenges, legal complexities, and economic disruptions \cite{kiritchenko2021confronting}.

\textbf{Mental Health Impact on an individual:} Toxicity emanating from online and AI systems can significantly harm an individual's mental health \cite{si2022so}. Exposure to online cyberbullying, harassment, and hate speech has been linked to increased anxiety, depression, stress, self-harm, and suicidal ideation, particularly among adolescents and marginalized communities \cite{twenge2020increases}. Excessive exposure to online toxic content can disrupt sleep patterns and elevate self-harm risks \cite{twenge2020increases}. The World Health Organization (WHO) has recognized problematic internet use as a growing public health concern, with online toxicity playing a significant role in reinforcing compulsive and harmful digital behaviors  \cite{kumar2018study, sandua2024double}

\textbf{Societal Impact:} Toxicity arising from Online and AI systems extends beyond individual harm, fueling societal issues like polarization, proliferation of mental health crises, and unabated spread of misinformation \cite{fan2024toxicity}. AI algorithms amplify ideological divisions by reinforcing pre-existing beliefs, making individuals more resistant to opposing views \cite{guess2023digital, biju2023self}. 

\textbf{Ethical Concerns:} The ethical implications afflicting online and AI systems are deeply rooted in questions of algorithmic responsibility, bias, and digital rights \cite{cheng2021socially}. AI-driven content moderation and recommendation systems often contain inherent biases that unequally impact certain demographic groups \cite{liu2024algorithmic}. Unregulated toxicity can distort public discourse, spreading misinformation, amplifying divisive narratives, and undermining trust in reliable information sources \cite{manheim2019artificial}.

\textbf{Legal and Regulatory Implications:} Governments and international bodies recognize the need for legal frameworks to regulate online toxicity and moderation of AI-driven content \cite{evangeline2025double}. The European Union’s Digital Services Act (DSA) and General Data Protection Regulation (GDPR) have established basic content moderation and user protection standards, setting a global precedent for the regulation of digital platforms \cite{bradford2023europe}. As digital communications continue to shape global discourse, addressing the challenges posed by toxicity in online platforms and AI systems is critical for ensuring a safer, more responsible, and equitable digital future \cite{stahl2021artificial}.

\textbf{Business and Technological Risks:}
Toxicity is more than a social, ethical, legal, or psychological issue; it is a silent killer \cite{george2023toxicity}. It is also a significant risk to the integrity, reputation, and long-term sustainability of online platforms, businesses, and AI-driven technologies \cite{polemi2024challenges}. When users encounter abusive content, misinformation, or targeted harassment, they are more likely to disengage affecting user retention, lower engagement rates, and a weakening sense of community \cite{throuvala2021perceived}.
The erosion of trust has far-reaching consequences. Toxicity can result in income loss, advertising reduction, and increased scrutiny and regulations \cite{beknazar2025toxic}. It damages brand credibility, discourages potential investors, and creates legal liabilities \cite{keller2018internet}. It challenges the credibility of automated systems. 
Addressing toxicity is not just an ethical obligation, but a crucial business strategy. Ensuring a safe and trustworthy environment is essential for long-term growth, user satisfaction, resilience, and technological reliability \cite{polemi2024challenges}.

\subsection{Stakeholders of the Toxicity Ecosystem} \label{sec_stakeholders} 
\begin{figure}[h]
\centering
\includegraphics[width=\linewidth]{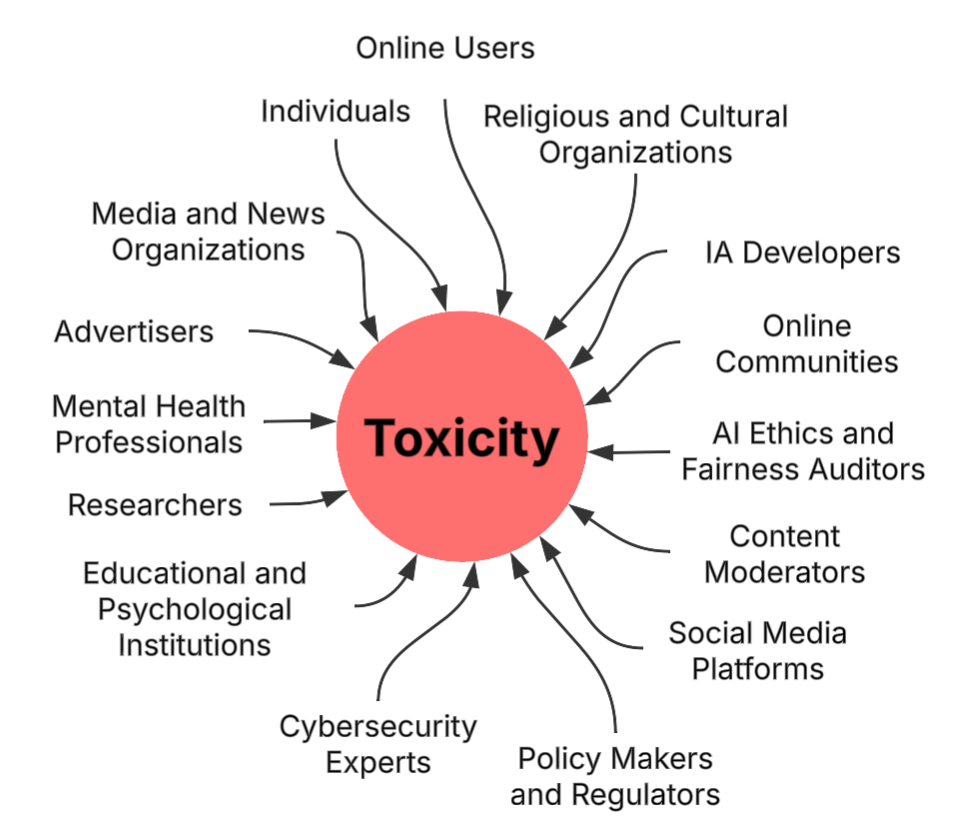}
\caption {Key Stakeholders in the ecosystem of online platforms and AI systems. Each stakeholder has a role in the generation, detection, moderation, and mitigation of toxic content.}
\label{fig:stakeholders}
\end{figure}

The prevalence of toxicity in online platforms and AI systems necessitates a collaborative effort among key stakeholders \cite{ognibene2023challenging}. Key stakeholders include individuals, platform owners, policymakers, AI developers, researchers, clinicians, businesses, media and news organizations, and educational and psychological testing institutions. Online users and individuals should be responsible for spreading and moderating toxic content, and SMPs should implement AI-driven moderation strategies \cite{siapera2022ai}. Online communities shape self-regulation norms, contributing to either the mitigation or amplification of toxicity \cite{zhang2024impact}. AI developers create automated moderation systems and toxicity control tools to enhance online safety and user experience \cite{zhang2022automatic}. Researchers study toxicity patterns and biases to improve detection mechanisms \cite{garg2023handling}. Policymakers and regulators implement and enforce regulations to hold platforms accountable for toxic content, fostering a safer digital environment \cite{flew2021regulating}. Businesses and advertisers prioritize brand safety by advocating for faithful content policies and responsible platform practices \cite{brown2020regulatory}. Media and news organizations influence public discourse on online platforms and toxicity of AI systems \cite{salminen2020topic}. Educational institutions promote digital literacy and study the psychological impact of toxic interactions  \cite{wardle2017information}. Figure \ref{fig:stakeholders} shows key stakeholders in the Online platforms and AI systems ecosystem.

\section{Systematic Review Methodology}

This section outlines the systematic methodology adopted to collect and analyze the literature on toxicity in online platforms and AI systems. Following established guidelines for systematic reviews, we employed the PRISMA framework ~\cite{moher2009preferred} to ensure reproducibility and transparency. The review process consisted of three main phases: keyword formulation, source selection, and document filtering.

\subsection{Keyword Formulation}

Given the interdisciplinary and evolving nature of toxicity in research on online platforms and AI systems, we designed a comprehensive keyword taxonomy spanning multiple conceptual and technical dimensions. Specifically, the search space was structured into six thematic categories, as shown in Table \ref{tab:keyword-categories}. 
This taxonomy ensured the retrieval of literature that addresses the phenomenon from various perspectives, including detection algorithms, social implications, model debiasing, explainability, and mitigation strategies. The complete set of keywords is presented in Table~\ref{tab:keyword-categories}.

\begin{table*}[ht]
\small
\centering
\resizebox{\textwidth}{!}{%
\begin{tabular}{p{0.135\textwidth}|p{0.15\textwidth}|p{0.225\textwidth}|p{0.17\textwidth}|p{0.15\textwidth}|p{0.13\textwidth}}
\hline
\textbf{Toxicity \& Hate Speech} & \textbf{Psychological \& Social Impact} & \textbf{Mitigation \& Detoxification} & \textbf{Detection Techniques} & \textbf{Bias and Fairness} & \textbf{Technical Terms} \\
\hline
Toxicity \newline Toxicity taxonomy \newline Implicit toxicity \newline Toxic language \newline Hate speech \newline Offensive language \newline Offensive language taxonomy \newline Abusive language \newline Cyberbullying
&
Psychological impacts of language toxicity \newline Mental health and online abuse \newline Effect of toxic language on users \newline Emotional impact of hate speech
&
Toxicity detection \newline Toxicity mitigation \newline Language detoxification \newline LLM detoxification \newline Hate speech moderation \newline Detoxifying large language models \newline Mitigation of toxicity in online platforms \newline Toxic content filtering \newline Bias mitigation in LLMs \newline Counterspeech detoxification
&
Toxicity detection \newline Offensive language detection \newline Hate speech detection \newline Toxicity detection using LLMs \newline Toxicity taxonomy \newline Bias detection
&
Bias in language models \newline Social bias detection in online platforms \newline Fairness in large language models \newline Bias in LLMs
&
Explainable AI in toxicity detection \newline Evaluation metrics for toxicity \newline Online platform evolution \\
\hline
\end{tabular}
}
\caption{Keyword categories used for the systematic literature review. Each column represents a thematic category, listing the search terms used to collect relevant literature on toxicity.}
\label{tab:keyword-categories}
\end{table*}

\subsection{Search Source}
We performed a comprehensive literature search across four prominent academic repositories: \textit{Google Scholar}, \textit{ACM Digital Library}, \textit{IEEE Xplore}, and the \textit{ACL Anthology}. These platforms were selected for their extensive coverage of computational linguistics, artificial intelligence, and natural language processing research.

Keyword queries were formulated using logical \texttt{OR} operators within each category and, where appropriate, combined using \texttt{AND} to ensure contextual specificity. For example:
\textit{("Toxicity" OR "Hate speech" OR "Cyberbullying") AND ("Toxicity detection" OR "LLM detoxification")}.
 The search covered works published between \textbf{2021} and \textbf{2025}, with an English-language restriction applied to narrow the scope and exclude studies focused on multilingual and low-resource language settings. Duplicate records were removed before the screening process.

\subsection{Document Screening and Inclusion Criteria}
 An initial corpus of \textbf{4,094 documents} was retrieved from the four selected databases. After removing duplicate records, a total of \textbf{2023 unique entries} remained for screening. Our team members independently reviewed the titles and abstracts of these entries to assess their relevance to the scope of this review. During this phase, studies were excluded if they were non-peer-reviewed, had fewer citations, were outside the scope of computational sciences, or lacked methodological or empirical relevance to toxicity in online platforms and AI systems.

Following the initial screening, \textbf{545 papers} were identified for full-text review. These articles were independently assessed by our team based on the following inclusion criteria: (1) the study addressed the detection or mitigation of hate speech, toxicity, offensive language, or cyberbullying; (2) it employed computational approaches relevant to online platforms or large language models (LLMs); and (3) it examined at least one key dimension such as bias mitigation, detoxification, or psychological impact. This structured evaluation resulted in the final inclusion of \textbf{247 peer-reviewed publications} for in-depth analysis. Among these, \textbf{232} focused on tasks related to toxicity in online platforms and AI systems.

Our methodology ensures a comprehensive and diverse corpus, representative of current research in toxicity detection, and provides a solid foundation for the analyses presented in subsequent sections.


\section{Modalities and Sources of Toxicity} \label{Sec_Taxonomy}

\subsection{Modalities of Toxicity} 
Toxicity in online platforms and AI systems exists in various modalities, including text, images, audio, video, and multi-modal. The severity of harm caused depends on how the content is consumed, shared, and interpreted. The speed and reach of digital communication have accelerated the growth of modern online platforms, also amplifying their effects \cite{skiba2024shadows}. Each mode of toxicity presents unique challenges in detection, moderation, and mitigation. This section provides an in-depth discussion of the different modes of toxicity, their impact, and why they are considered toxic in online platforms and AI systems.

\subsubsection{Text-based Toxicity}
Text-based toxicity is one of the most widespread and studied modalities of toxic and harmful content. The ability to quickly generate, modify, and distribute text across multiple platforms makes it a powerful medium for constructive and destructive discourse \cite{keen2011cult}. Textual toxicity is particularly harmful due to its scalability. One of the primary reasons text toxicity is harmful is its direct psychological impact \cite{scheuerman2021framework, saha2019prevalence}. For instance, harassment, hate speech, and aggressive language can cause stress, anxiety, and depression \cite{bilewicz2020hate}. Text-based toxicity is aggravated by anonymity and platform structures that enable rapid dissemination \cite{singh2022technical}. 

\subsubsection{Image-based Toxicity}
Toxic images may display not only explicit content, but also spread misinformation and visual hate speech, and become viral \cite{pandiani2024toxic}. Images have an extraordinary power because they are processed faster by the human brain than text and are often more emotionally striking \cite{greenfield2015mind}. A significant challenge with image-based toxicity is its ability to bypass traditional moderation methods \cite{li2025t2isafety}. Explicit or violent images are relatively easy to detect. However, more indirect forms of image toxicity, such as hate symbols and deepfakes, are difficult to recognize \cite{aprin2024supporting}. Detecting these forms of toxic image content requires advanced mechanisms. Image toxicity is particularly harmful due to its viral nature. 

Manipulated images, offensive visual content, and graphic violence can spread rapidly across online platforms \cite{duncombe2020social}. Such widespread dissemination can have significant consequences in the real world, affecting social behavior, politics, and legal outcomes. Misleading toxic images can provoke violence and fuel political and social conflicts \cite{gamson1992media}. Toxic images generated by AI, such as fake content, create a more significant challenge. Advances in generative AI enable the creation of hyper-realistic images, which can be used for offensive purposes, including generation of misinformation \cite{wang2023synthetic}. Such content can seriously harm individuals and institutions. The legal and ethical implications of AI-generated imagery remain a topic of ongoing debate worldwide \cite{florindi2024ethical}.

\subsubsection{Audio-based Toxicity}
Audio-based toxicity is becoming widespread with the rise of voice assistants, social audio platforms, and AI-generated speech technologies \cite{yu2023antifake}. Unlike text and images, audio-based toxicity presents a significant challenge due to its temporal nature \cite{yousefi2021audio}. The harm caused by toxic audio content often arises from the emotional intensity conveyed through spoken words \cite{soni2018see}. Elements like tone, pitch, and emphasis can significantly shape the meaning of a statement. As a result, spoken or audio toxicity can sometimes be more damaging than written words. Detecting and moderating audio toxicity is difficult due to linguistic and cultural diversity. Detection and moderation become more complex when it is processed in real-time. Slurs, coded language, and harmful audio toxicities may not always appear explicitly offensive. However, their impact can vary according to the speaker's intent, tone, and contextual nuances.

Another emerging challenge is the use of AI-generated voices for toxicity \cite{desai2024gen}. Deepfake audio can realistically mimic real people, enabling impersonation, fraud, and targeted harassment. This technology has been exploited to create fake audio clips of political figures, spread misinformation, and manipulate public perception \cite{desai2024gen}). The consequences of synthetic audio manipulation extend across politics, journalism, and cybersecurity. Real-time audio toxicity, such as harassment in various communities in online communications, has been especially difficult to address \cite{blackwell2019harassment}. Toxic audio messages delivered in situations when one is engaged in activities such as driving can also be life-threatening. Live voice interactions are challenging to monitor and moderate effectively. In addition, victims often struggle to report toxic behavior, as conversations disappear once they end, leaving no evidence \cite{saarinen2017toxic}.

\subsubsection{Video-based Toxicity}
Toxicity delivered via video is multidimensional combining text, audio, and visual elements. As a result, it is very difficult for online platforms and AI systems to recognize and mitigate. Toxic videos can contain explicit content, hate speech, misinformation, or violent material \cite{arora2023detecting}. Due to its immersive nature, video content can be more powerful and influential than other modalities \cite{tuong2014videos}. This increases the impact of toxic video content, making moderation even more challenging \cite{gongane2022detection}. The main concern with video toxicity is its ability to manipulate emotions and perceptions on a large scale. Videos are highly engaging and easily shareable, often reaching a broad audience before moderation \cite{alshami2024smart}. Toxic content in videos has been linked to harassment campaigns and the spread of conspiratorial thinking. This makes video-based toxicity very dangerous and difficult to control \cite{maity2024toxvi}. The rise of AI-generated deepfake videos has made it easy to incorporate harmful content in seemingly realistic videos. Deepfakes can be highly realistic synthetic videos that can misrepresent individuals and fake events \cite{hancock2021social}. Deepfake-powered video toxicity presents serious threats to political stability, personal integrity, and legal systems. As deepfakes become more sophisticated, detecting and preventing video-based toxicity becomes a significant challenge \cite{nasri2018investigation}.

\subsubsection{Multi-Modal Based Toxicity}
Multi-modal toxicity involves blending of text, image, audio, and video to create a more complex and often more dangerous form of toxic content \cite{maity2024toxvidlm}. It leverages multiple forms of inputs, making toxic and harmful content more engaging and impactful. The rise of toxic memes, which blend text and imagery to create offensive or toxic content is a perfect example. Memes spread toxicity effectively using humor and sharing to broadcast toxic content \cite{kiela2020hateful}. Due to the interplay of different content types, multi-modal toxicity is challenging to detect and moderate \cite{liu2024arondight}. Traditional AI moderation tools struggle to analyze cross-modal interactions and toxicity \cite{nedungadi2025ai}.

\subsection{Information-based toxicity} \label{Info-based-Toxicity}
Online platforms facilitate businesses, education, gaming, banking and financial services, healthcare, transportation, news media, legal, public services, and many more. The content on these platforms is either user-generated or AI-generated. It includes articles, reports, notifications, reviews, forms, chatbots, and many other formats. 

When the content is generated by the subscriber or the user of a particular platform, it is referred to as user-generated content \cite{wyrwoll2014usergencontent}. Regardless of mode, such content in any mode is generated with or without the help of AI tools or by hand. The content that uses Generative AI models \cite{cao2025survey} is usually referred to as AI-generated content, discussed in detail in Section \ref{Sec_GenAI}. Such content has the potential to host at least one or all of the forms of toxicity. Both user-generated and AI-generated content may host various fine-grained toxicities, explained in detail in Section \ref{Sec_Fine}.


\subsection{Interaction-based Toxicity} Various forms of digital communication like emails, messenger, and social media permeate our lives, providing information pertaining to news, education, business, healthcare, social engagements, government engagements, and so on. These enhance the interaction among people and entities. On various occasions, they host toxicity impacting social or physical well-being. 

\subsubsection{Social Media Platforms} 
Social Media Platforms\cite{dhingra2019historical}, as the name suggests, are meant to bring people together. People share opinions, moments, pictures, videos, and stories informally. Over the last 2 decades, SMPs have grown to be a multi-billion-dollar business. SMPs have added various features like followers, reactions, comments, replies, streaming services, advertisements, feeds, and political and social influencers, to mention the least. The primary source of revenue for SMPs is user engagement, i.e., how long users are glued to it. The algorithms implemented track users' activities. Based on users' reactions, a platform keeps flooding their accounts with relevant feeds, which may impact users psychologically. Arguably, users get addicted to SMP and disconnected from mainstream communications. The 2020 Netflix Documentary \textit{The Social Dilemma}, directed by Jeff Orlowski, explains this phenomenon, in particular how it impacts youth and adults, and how it moves people to extremism without realization. The main attributes of social media that engender toxicity in SMP are the following \cite{khapre2025mitigation}.
\begin{itemize}
    \item \textbf{Engagement:} Users engage in an SMP either by post or feed. They can comment, reply, and share posts or feeds. These acts, in turn, may exhibit their own biases \cite{cheng2024like, smp_UE_shahbaznezhad2021role}
    
    \item \textbf{Reactions:} Users react to the posts and feeds using emojis \cite{eberl2020s, lin2015emotional, cheng2024like}. Negative reactions may introduce toxicity, depending on the context. Another potent form of reaction is to un-follow or un-friend a user. A reduction in the number of followers or friends may impact upon an individual's self-esteem or affect in other ways. Negative reactions may signal criticism, disapproval, or denial of relationships among the users, potentially impacting their well being.
    
    \item \textbf{Algorithmic Bias:} \cite{qiu2024socialmedia, thomas2022overview}
    Two types of algorithms employed in an SMP \textit{promote} user engagement in the addictive content and for \textit{revenue} generation from advertisement feeds. The algorithms prioritize emotionally provocative or controversial content to maximize user interaction \cite{zollo2018misinformation, vosoughi2018spread}.
    
\end{itemize}
SMPs are afflicted with fine-grained toxicities discussed in Section \ref{Sec_Fine}. Although knowingly or unknowingly, an SMP may host a lot of these toxicities, with concomitant negative impacts on society and human psychology \cite{wyrwoll2014usergencontent, vaidya2020empirical, shu2020fakenewsnet, gao2020public, duncombe2020social, smp_UE_shahbaznezhad2021role, throuvala2021perceived, gongane2022detection, ognibene2023challenging, sharma2023security, antypas2023TweetHate, aprin2024supporting, zhang2024impact, dangsawang2024machineSmuggledGoods, roy2024analysisIllegalWildlife, ray2024cyberbullying, ali2024youth, przybyla2024know, fan2024toxicity, qiu2024socialmedia, hall2025loneliness, ali2025impact, beknazar2025toxic, khapre2025mitigation}. SMPs are popular in society for a lot of their positive contributions. 

\subsection{Generative AI} \label{Sec_GenAI}

Generative AI has revolutionized digital content creation, enabling the rapid synthesis of text \cite{shakil2024abstractive}, images \cite{archana2024analysis}, audio \cite{mitra2025music}, and video \cite{zhou2024survey}. Large Language Models (LLMs) such as OpenAI’s GPT \cite{shakil2024evaluating}, Google’s Gemini \cite{imran2024google}, and Meta’s LLaMA \cite{li2025comparative} generate human-like text by leveraging vast datasets. However, these models are also prone to producing toxic content, including hate speech, misinformation, and offensive material, either unintentionally or as a byproduct of biased training data \cite{gehman2020realtoxicityprompts, vidgen2021learning}. The generative capability of AI introduces significant challenges in online toxicity management, as it can reinforce harmful biases, amplify misinformation, and propagate toxic content at scale \cite{sap2020social, mathew2021hatexplain}. 

Generative AI interacts with online toxicity in several ways. First, it learns from unfiltered internet data, which often contains implicit \cite{wen2023unveiling} and explicit toxicity \cite{gunturi2023toxvis}, leading to biased or harmful outputs \cite{wei2025addressing}. For example, LLMs trained on datasets with racial or gender biases may inadvertently perpetuate these biases in their outputs, even when explicitly instructed to avoid toxic content \cite{bender2021dangers}. 
Second, generative AI enhances engagement by generating content that aligns with user preferences, which can sometimes exacerbate toxic behaviors \cite{garg2023handling}. It is particularly true in SMP, where algorithms prioritize emotionally provocative or controversial content to maximize user interaction \cite{zollo2018misinformation, vosoughi2018spread}. 
Third, generative AI hallucinates, ie, it produces misleading or fabricated information, contributing to misinformation and digital harm \cite{huang2025survey}. Hallucinations can lead to the spread of false narratives, which are often more engaging and emotionally charged than factual content, making them more likely to go viral \cite{shakil2024utilizing}.

Addressing these issues requires a nuanced understanding of how AI-generated toxicity manifests and impacts online discourse. For instance, the ethical implications of generative AI in content creation are profound, including questions about accountability, transparency, and the potential for misuse in spreading disinformation or hate speech \cite{dubber2020oxford}. Moreover, the challenges of detecting and mitigating toxicity in multi-modal content (e.g., toxic memes, deepfake videos) further complicate the landscape, as traditional text-based moderation tools are often insufficient to address these emerging threats \cite{desai2024gen}.

\subsubsection{Engagement}
Engagement in digital platforms is driven by algorithms that maximize user interaction, often by prioritizing emotionally provocative or controversial content \cite{zollo2018misinformation, vosoughi2018spread}. Generative AI plays a significant role in shaping these engagement patterns, as its outputs are designed to be compelling and personalized. While AI-generated content fosters meaningful interactions, it also risks reinforcing toxic discourse through both implicit and explicit engagement. This dual nature of engagement poses significant challenges for online platforms, as it can amplify harmful behaviors while simultaneously driving user retention and revenue \cite{selvakumar2025balancing}.

\textbf{Implicit} engagement refers to AI’s ability to subtly shape user interactions without overt toxicity. Generative AI models reinforce biases by personalizing content recommendations based on user behavior, which can lead to echo chambers and filter bubbles \cite{jawad2024investigating, siapera2022ai}. These echo chambers intensify ideological divides, making users more resistant to opposing views and indirectly perpetuating toxicity \cite{mathew2021hatexplain}. For example, AI systems that recommend content based on user preferences may inadvertently reinforce harmful stereotypes or prejudiced language structures, even when the content itself is not explicitly toxic \cite{garg2023handling, zellers2019defending}.


\textbf{Explicit} engagement occurs when AI-generated content directly propagates toxic discourse, often through misinformation, hate speech, or inflammatory language \cite{fitzgerald2020distinct, gambin2024deepfakes, mersha2024ethio}. AI systems trained on unfiltered datasets are prone to generating harmful content that fuels online harassment, cyberbullying, and hate speech \cite{ali2024youth}. For instance, toxic memes, deepfake videos, and offensive chatbot responses illustrate how generative AI can contribute to explicit online toxicity \cite{desai2024gen}. These forms of toxicity are particularly dangerous because they are designed to provoke strong emotional reactions, making them highly shareable and difficult to moderate \cite{lehtimaki2024navigating}.

The engagement-driven nature of online platforms exacerbates explicit toxicity by rewarding high-interaction content. Studies indicate that AI-generated misinformation spreads significantly faster than factual content due to its provocative and emotionally charged nature \cite{vosoughi2018spread}. This presents a critical challenge for content moderation, as automated AI models often fail to distinguish between engaging content and harmful narratives \cite{siapera2022ai}. For example, AI systems may prioritize content that generates high engagement, even if it contains hate speech or misinformation, due to the lack of robust mechanisms for detecting and filtering such content in real time \cite{wang2023survey}.

Addressing explicit AI-driven toxicity requires improved content moderation strategies, enhanced model transparency, and the integration of bias-mitigation techniques such as adversarial training and reinforcement learning with human feedback \cite{liu2023transforming}. Additionally, platforms must adopt a multi-stakeholder approach, involving policymakers, researchers, and civil society organizations, to develop ethical guidelines and regulatory frameworks that balance user engagement with online safety \cite{dubber2020oxford}.

\subsubsection{Hallucination}

Hallucination in generative AI refers to the phenomenon where models generate incorrect, misleading, or fabricated content, often presented as factual information \cite{ji2023survey, shakil2024utilizing}. This issue is particularly problematic in toxicity management, as hallucinated outputs can contribute to the spread of misinformation, hate speech, and harmful stereotypes \cite{tonmoy2024comprehensive}. Hallucinations arise from several factors, including gaps in training data, overgeneralization, and the probabilistic nature of language model generation \cite{guerreiro2023hallucinations}. When AI systems lack sufficient context or rely on incomplete datasets, they generate content that may appear plausible but is factually incorrect \cite{huang2024calibrating}.

The impact of AI hallucinations extends beyond misinformation, influencing public opinion and exacerbating social biases. For instance, AI-generated content can reflect and amplify racial, gender, and political biases present in the training corpus, perpetuating toxic discourse \cite{fang2024bias}. In the context of large-scale SMP, hallucinated toxicity can rapidly spread, leading to real-world consequences such as increased polarization, targeted harassment, and even violence \cite{latif2025hallucinations}. For example, AI-generated fake news or deepfake videos can manipulate public perception, erode trust in institutions, and destabilize democratic processes \cite{wang2023synthetic}.

Efforts to mitigate hallucinations in generative AI focus on refining training methodologies and improving model oversight. Techniques such as knowledge-grounded generation, fact-checking mechanisms, and adversarial training have shown promise in reducing AI hallucinations \cite{yu2024cause}. Reinforcement learning with human feedback further enhances model reliability by aligning outputs with human values and factual accuracy \cite{ouyang2022training}. However, balancing creativity with factual correctness remains a challenge, particularly for models generating diverse and context-dependent content. For instance, while reinforcement learning with human feedback can reduce hallucinations, it may also limit the model's ability to generate creative or contextually rich responses, highlighting the need for a nuanced approach to hallucination mitigation \cite{yaprak2024generative}.

Moreover, addressing hallucinations requires a multi-faceted approach that combines technical solutions with ethical and regulatory frameworks. For example, platforms must implement robust content moderation systems to detect and remove hallucinated content, while policymakers must establish guidelines to hold AI developers accountable for the societal impact of their systems \cite{bertoncini2023ethical}. Collaborative efforts between researchers, developers, and policymakers are essential to ensure that generative AI systems are both innovative and responsible \cite{kim2024unveiling}.


\section{Fine-Grained Toxicities} \label{Sec_Fine}
\begin{figure}[h!]
    \centering
    \includegraphics[width=1 \linewidth]{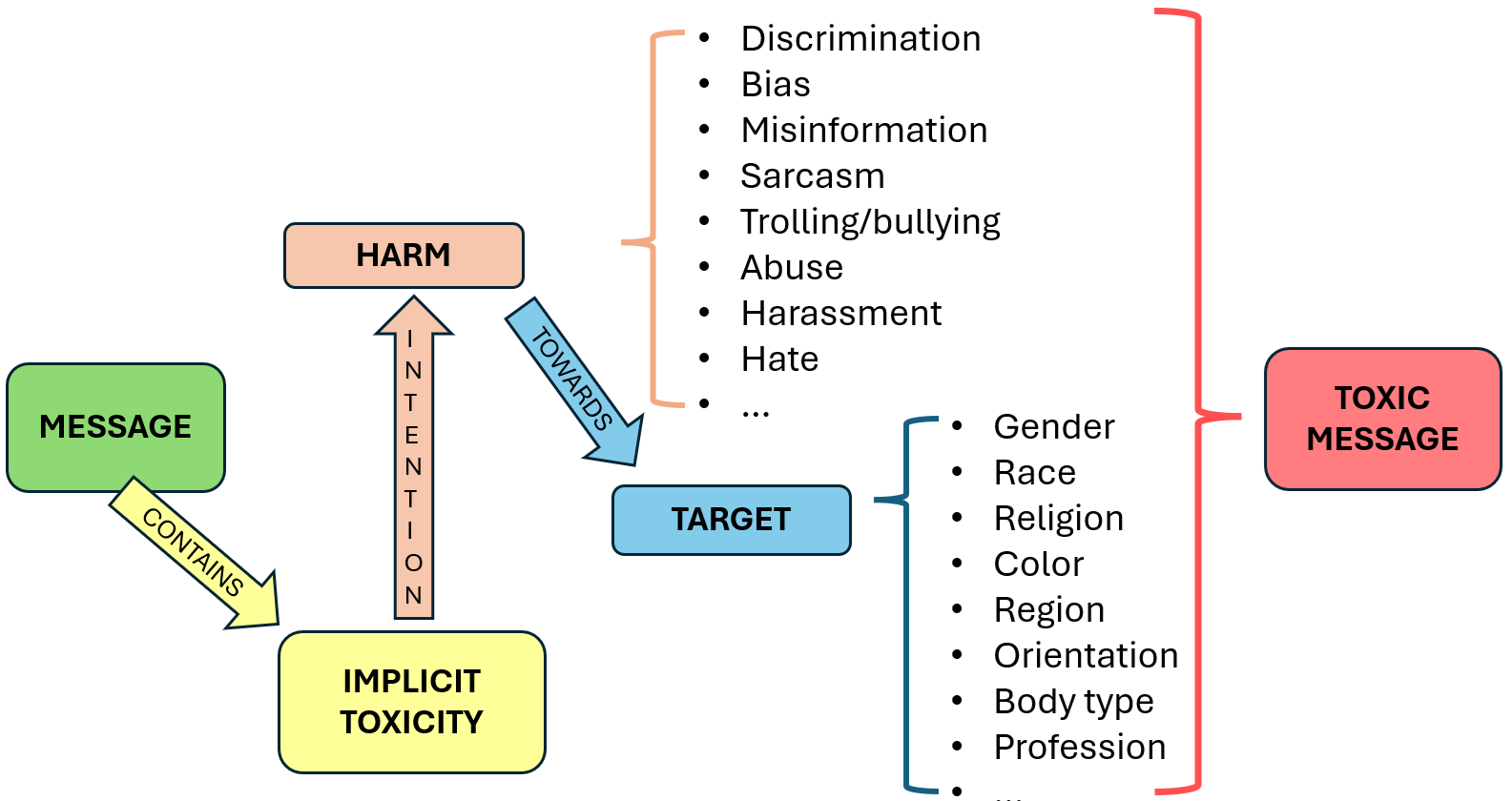}
    \caption{Implicit Toxicity: The message seems benign or normal, but based on the context, the reader or viewer, or the environment, it intends to harm a particular target. Such messages have implicit toxicity.}
    \label{fig:Implicit}
\end{figure}

\begin{figure}[h!]
    \centering
    \includegraphics[width=.9 \linewidth]{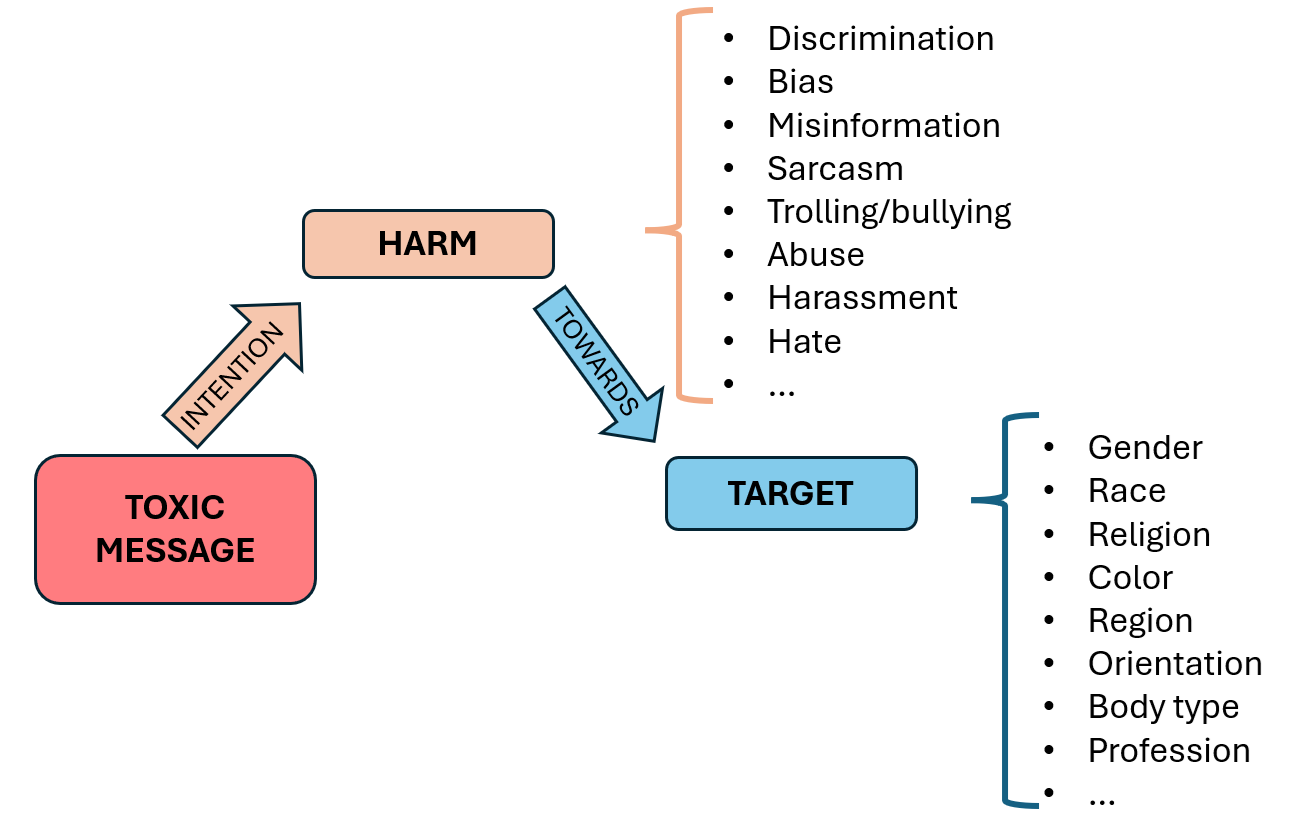}
    \caption{Explicit Toxicity: The explicit toxic message intends to harm a particular target, which can be easily understood as toxic. Its implied meaning and actual content have the same meaning. Some explicit toxic messages that do not have a target are prevalent in the form of profanity, sarcasm, and other forms of offensive language.}
    \label{fig:Explicit}
\end{figure}

The main types of toxicities are implicit toxicity, as depicted in Figure \ref{fig:Implicit}, explicit toxicity, as depicted in Figure \ref{fig:Explicit}, and fallacy. Explicit toxicity is also called fallacy by some. Fallacy \cite{helwe2024mafalda} refers to misinformation, disinformation, and fake news. Often, toxicity has intentions of causing harm and is targeted towards a particular target group \cite{mersha2024ethio}. There is another important type of \textit{toxicity without a target} which could be both implicit or explicit, containing profanity, negative, derogatory, disrespectful, and offensive language \cite{pachinger2023toward}. Before 2021, this type was the focus of study and research as detailed by van Aken et al. \cite{van2018challenges}. Despot et al. highlighted that while analyzing implicit toxicity with figures of speech like metaphor, explicit toxic content without a target tends to be ignored \cite{vstrkaljsomewhere, despot2023somewhere}. They have presented a typology of implicit toxicity such that explicit toxicity gets detected and not ignored. It includes caps lock and repetition as aggressive speech; irony, metaphor, circumlocution, hyperbole, rhetorical questions, (re)interpretation, euphemism, simile, contrast, and name-calling are fine-grained in discrediting, insulting, dehumanization, derogation, and discriminatory speech.

Pachinger et al. \cite{pachinger2023toward} have analyzed various research and analyzed a granular enumeration of harms caused by toxicity, including threats, rudeness, mockery, name-calling, obscenity, pejorative content, aggressiveness, profanity, insult, discrimination, discrediting, irony, sarcasm, derogation, hostility, hatefulness, vulgarity, and criminality. 

Wang et al. \cite{wang2024Donotanswer} have come up with a taxonomy of LLM risks containing three levels with an extensive fine-grain at the third level. They have highlighted that the harms of toxicity may also be a result of illegal trade, and illegal activities such as abuse, cruelty, exploitation, intimidation, cyberbullying, harassment, baiting, stalking, and defamation of vulnerable individuals. Harm could also arise from the endorsement of unsafe items or practices, and the spread of information hazards such as personally identifiable information pertaining to life, health, finance, education, biometrics of an individual, confidential, classified, legal, and sensitive in nature. Such information may cause cybersecurity vulnerabilities, theft of technical data and internal communication, misinformation, disinformation, fake news, rumors, propaganda, misinterpretation, and unreliable experts giving legal, medical, and financial advice. Fine-grained target groups include those targeted for age, disability, gender, sexual orientation, race, ethnicity, religion, region, color, body type, language, class, nationality, profession, and others. 
B{\k{a}}czkowska et al. \cite{bkaczkowska2025implicit} presented an implicitness model comprising indirectness in rhetorical questions, figures of speech like simile, metaphor, and irony (sarcasm), exaggerations in forms of over- or understatement; resulting in hostile, hateful, racist, homophobic, discredit, threat, misogynic and vulgarism. Guest et al. \cite{guest2021Misogyny} presented a detailed misogyny taxonomy based on online misogyny literature, mainly on social media. They have classified online misogyny into 4 major classes of derogatory, pejorative, treatment, and personal attacks; the pejorative and personal attacks in their taxonomy seem explicit toxicity; the other two have further fine-grained toxicities, like derogatory, which have intellectual and moral inferiority, sexual and physical limitations, and others. Treatment has threatening language involving physical and sexual violence, invasion of privacy, and disrespectful actions involving controlling, manipulation, seduction, conquest, and others.

Korre et al. \cite{korre2024hate} have introduced a compelling perspective of the legal framework in understanding hate speech, emphasizing its prosecutable nature and significant variation across countries. They have highlighted three legal approaches: content-, intent-, and harm-based, having some target being impacted. The content-based approach has direct content to offend certain people in society; intent-based is communication to its audience to invoke hatred towards minorities or other hateful intentions, evidently not legitimate; and harm-based induces lower self-esteem, physiological, emotional, and psychological conditions, socio-economic distress, and societal withdrawal symptoms to its victims.

\section{Convergence with Human Psychology} \label{Sec_Psycho}
In human psychology, toxicity is usually referred to as \textit{incivility,  offensive behavior} \cite{lewandowska2023integrated}, \textit{flaming}, and digital communication as \textit{computer-mediated communication} \cite{lee2005behavioral, bansal2012classification}. Flaming (aka Toxicity) can further be considered as direct or indirect, and straight or satirical \cite{bansal2012classification}. Ali et al. \cite{ali2025impact} demonstrated the relation of social media platform addiction with impacts like depression, anxiety, bad interpersonal relationships, hopelessness, guilt, sadness, lack of enthusiasm, sleep, and compulsive disorders, and called for mitigation strategies in digital platforms. See Section \ref{Sec_Whymatters} for mental health impact. Papageorgiou et al. \cite{papageorgiou2024survey} highlighted the need for safeguarding information integrity to ensure societal trust in fallacy detection. Pennycook et al. \cite{pennycook2021psychology} investigated the psychology behind fallacy and why it spreads faster than real news. 

\subsection{Biases in LLMs}

The bias embedded during training significantly affects the development of fair AI systems. A growing number of studies \cite{rao2023can} indicate that LLMs exhibit reasoning and self-improvement abilities that resemble human cognitive processes, raising the possibility that these models may develop virtual personalities and psychological characteristics \cite{pan2023llms}. In particular, gender- and race-based discriminatory content is often introduced during training \cite{cabrera2023ethical}, which can embed harmful biases into the model’s responses and behavior. These biases can manifest in stereotyping, preference for dominant social norms, or assumptions based on patterns that do not reflect the complexity of real-world experiences \cite{ferrara2023should}. The high parameterization of modern LLMs, while enhancing capacity, introduces multidimensional interactions that are difficult to predict and manage \cite{trenk2024text}.
Bai et al. \cite{bai2024measuring} reported that the most substantial biases are found in categories related to race, followed by gender, health, and religion. One illustrative example of gender bias is the reinforcement of exclusionary language norms, such as referring to “both genders,” which implicitly excludes non-binary identities \cite{bender2021dangers}. 
Despite increased attention to these challenges, much of the current bias research in LLMs falls short in clearly identifying who is harmed, why the behavior is harmful, and how it reflects and reinforces social hierarchies \cite{blodgett2020language}. 
While all LLMs exhibit some level of bias, there are significant variations among them. Models with more parameters, such as GPT-4, GPT-3.5-Turbo, Claude3-Opus, and Claude3-Sonnet, as well as LLaMA2Chat-70B and 13B, tend to display more pronounced implicit biases. In contrast, smaller models like LLaMA2Chat-7B and Alpaca7 B show substantially lower bias levels \cite{kumar2024investigating}.

\subsection{Discrimination in LLMs}

Discrimination in LLMs can manifest in both direct and indirect forms. Direct discrimination involves overtly unequal treatment based on group membership. For instance, when LLM-powered resume-screening tools reinforce existing hiring inequities \cite{ferrara2023should}. Indirect discrimination occurs when model outputs, while seemingly neutral, rely on proxy variables that disproportionately disadvantage certain groups. This can be seen in LLM-assisted healthcare systems, where demographic proxies contribute to inequities in patient care \cite{ferrara2023should}.
Many approaches, for example, rely on the implicit assumption that model outputs should be independent of social group identifiers, yet they often fail to explicitly articulate the underlying social or ethical values that support this assumption. Furthermore, inconsistency in how bias is defined and a lack of meaningful engagement with structural power dynamics hinder the creation of robust and equitable solutions \cite{blodgett2020language}.

\subsection {Personality Types and Toxicity}
Wang et al. \cite{wang2025exploring} showed in their experiments using the HEXACO framework on three LLMs, that the low levels of  sensitivity of these six personality traits of \textit{Honesty-Humility, Emotionality, eXtraversion, Agreeableness, Conscientiousness,} and \textit{Openness to Experience}, produce higher levels of toxicity and bias in their responses. Garcia et al. \cite{serapio2023personality} through experiments have shown that the LLMs have a synthetic personality and can be fine-tuned to mimic desired personality traits. Pam et al. \cite{pan2023llms} used another psychological test MBTI (Myers-Briggs Type Indicator), with dichotomies of extraversion, introversion, sensing, thinking, intuition, feeling, judging, and perceiving. They have shown that LLMs innately possess some personality traits.
In a similar study, Jiang et al. \cite{jiang2024personallm} used \textit{Big Five Inventory} test on LLMs with personality profiles, and their self-reported evaluations were consistent with the LLMs' personality profiles. All these psychological evaluations of LLMs' personality traits provide a mitigation strategy that can be built into the LLMs by fine-tuning them with higher sensitivity, reducing bias and toxicity in LLM-generated content. At the same time, one should be careful during LLMs training, so that it avoids less sensitive personality traits which might otherwise make it generate toxic contents.

\subsection {Issue-Based Discussions and Political Polarization}
Although digital platforms allow open discourse among diverse populations, they are also breeding grounds for toxicity in politically charged conversations. Gao et al. \cite{gao2024crisis} analyzed over 40 million Reddit comments and found that incivility, including vulgarity, aspersion, and name-calling, is more prevalent in politically biased forums. People are more likely to respond uncivilly to those with opposing views. Moreover, toxic comment threads tend to attract more participation but also escalate in hostility. Exposure to personal attacks often leads users to disengage from political spaces. If LLMs are trained on such content, they may inadvertently learn and replicate these patterns, reflecting partisan toxicity and asymmetrical ideological bias. For instance, certain forms of incivility may be overrepresented by specific political groups, which in turn can influence how models respond to politically sensitive prompts.
Gao et al. \cite{gao2024crisis} also found that both left-leaning and right-leaning individuals became more uncivil when engaging with opposing views, though right-leaning users exhibited higher levels of all incivility subtypes except third-party attacks. The most prominent difference—interpersonal name-calling—was 177 percent higher among right-leaning users.
In a related study, Rozado \cite{rozado2023political} administered 15 political orientation assessments to ChatGPT and reported a tendency for the model to favor left-leaning perspectives. However, this finding should be interpreted cautiously due to the limited number of tests and the exploratory nature of the analysis.

In conclusion, LLMs not only mirror the toxic, biased, and polarized content present in online human communication but also risk reinforcing these tendencies when not properly constrained. Understanding these interactions is crucial for developing more equitable, responsible, and psychologically aware AI systems.


\section{Toxicity Mitigation} \label{Sec_Mitigation}
\begin{figure*}[h!]
    \centering
    \includegraphics[width=1 \linewidth]{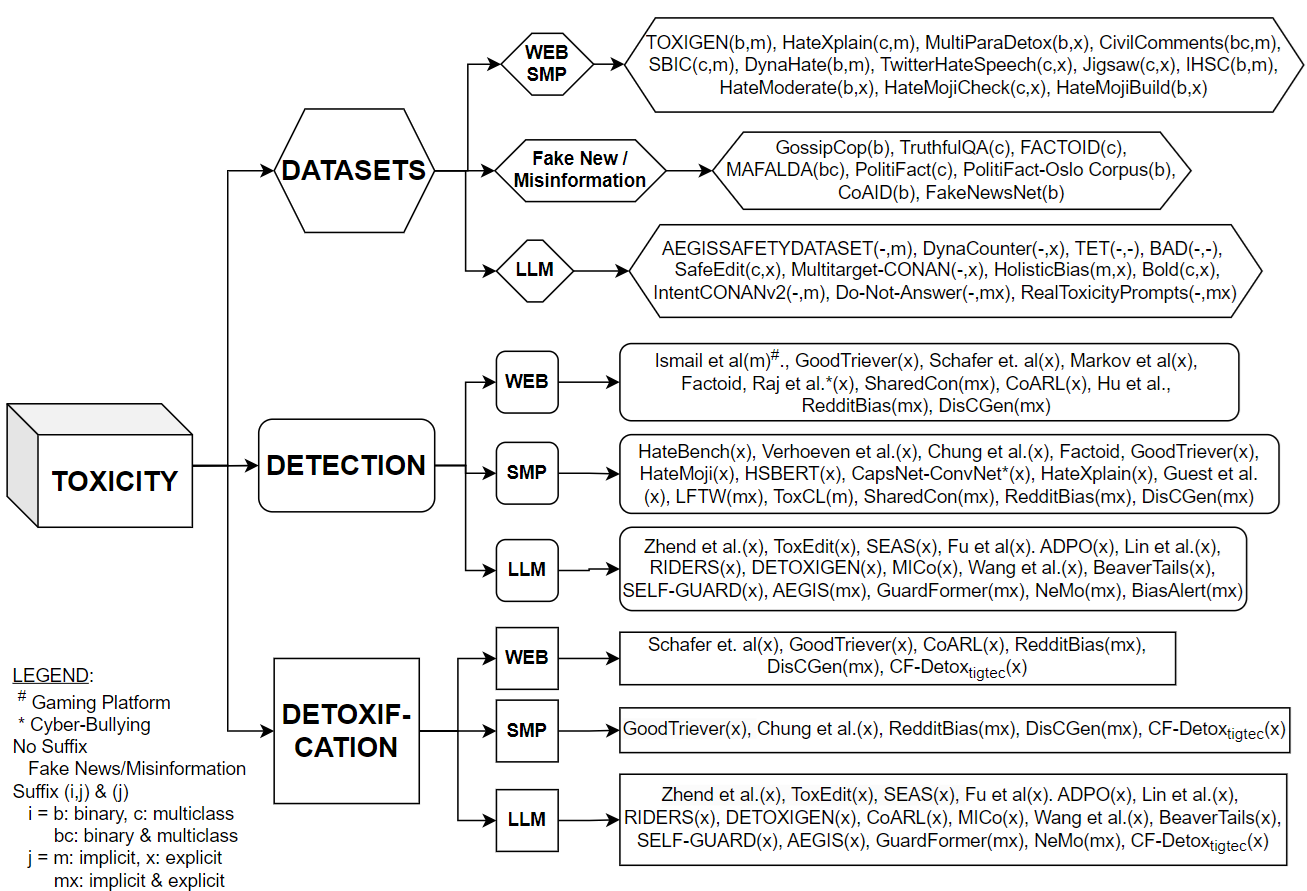}
    \caption{Summary of Toxicity Mitigation mechanism, which includes dataset, detection, and/or detoxification mechanisms. Their references are listed in the Table \ref{tab:Toxicity_Datasets} and Table \ref{tab:detection}}
    \label{fig:T-mit}
\end{figure*}

\begin{figure}
    \centering
    \includegraphics[width=1\linewidth]{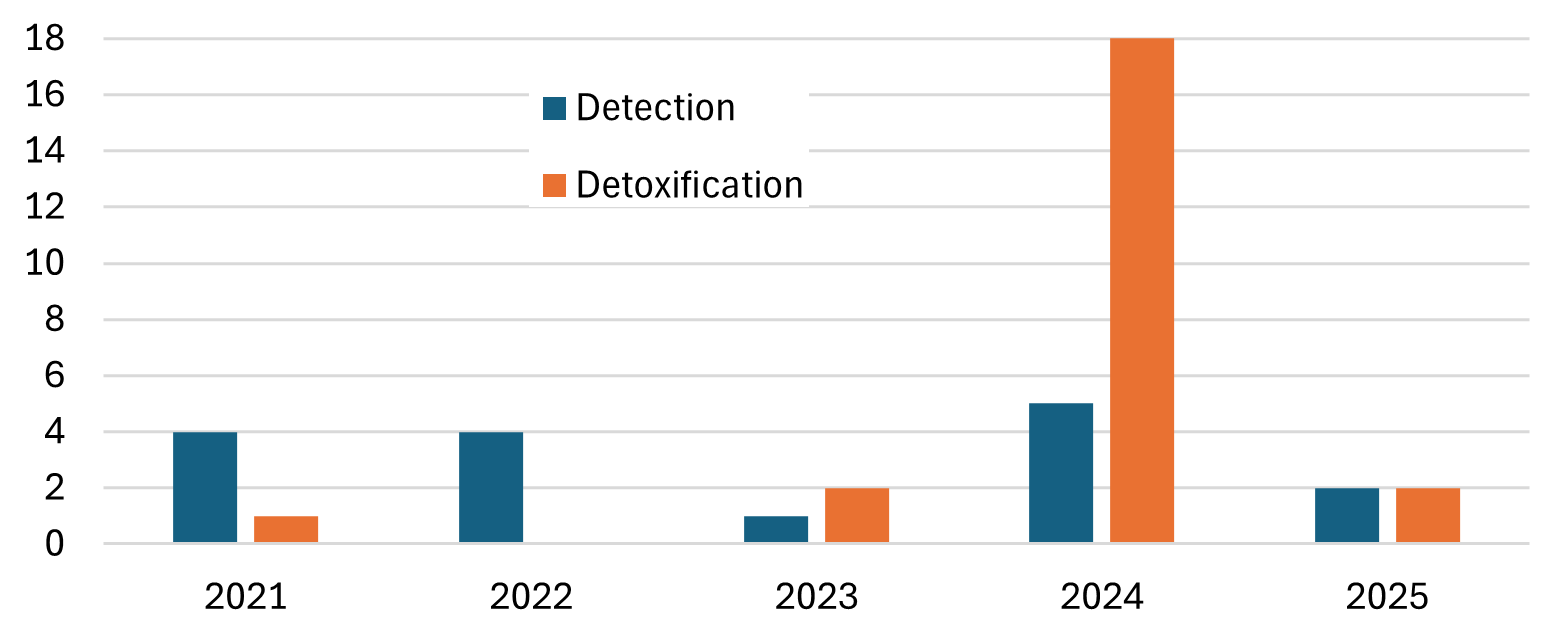}
    \caption{Toxicity mitigation research since 2021. Y-axis is the number of research papers published.}
    \label{fig:mitigationByYear}
\end{figure}

Toxicity mitigation mechanism requires a dataset, detections, and detoxification mechanisms (framework, model, or techniques). All the surveyed literature is summarized in Figure \ref{fig:T-mit} based on broadly classified domains as Web or SMP, Fake News or Misinformation, and LLM. Most of the models developed or proposed can be tested in at least one of these domains. Web or SMP toxicity requires a dataset of text sequences taken from SMP, gaming, healthcare, education, or other web-based platforms for detection models, and counterspeech for moderation or detoxification. Fake news or misinformation requires a fallacy-based strategic detection model, where detoxification is quite challenging. The fallacy datasets are created using Gossipcop and Politifact web databases. And LLMs are the model itself and require design modification within the model to allow detection and detoxification. Toxicity datasets for LLMs are datasets of question-answer pairs. In Figure \ref{fig:T-mit}, each listed dataset is suffixed with two parameters in parentheses, \textit{'i'} and \textit{'j'}. Each listed detection and detoxification method has a suffix in parentheses \textit{'j'}. Where \textit{'i'} is either binary \textit{'b'} or multi-class \textit{'c'} or both \textit{'bc'}. The recent momentum in research papers published by year is shown in Figure \ref{fig:mitigationByYear}. The toxicity mitigation research has shown momentum in 2024 using LLMs as detailed in Table \ref{tab:detection}.

\subsection{Toxicity Datasets}
\begin{figure}
    \centering
    \includegraphics[width=1\linewidth]{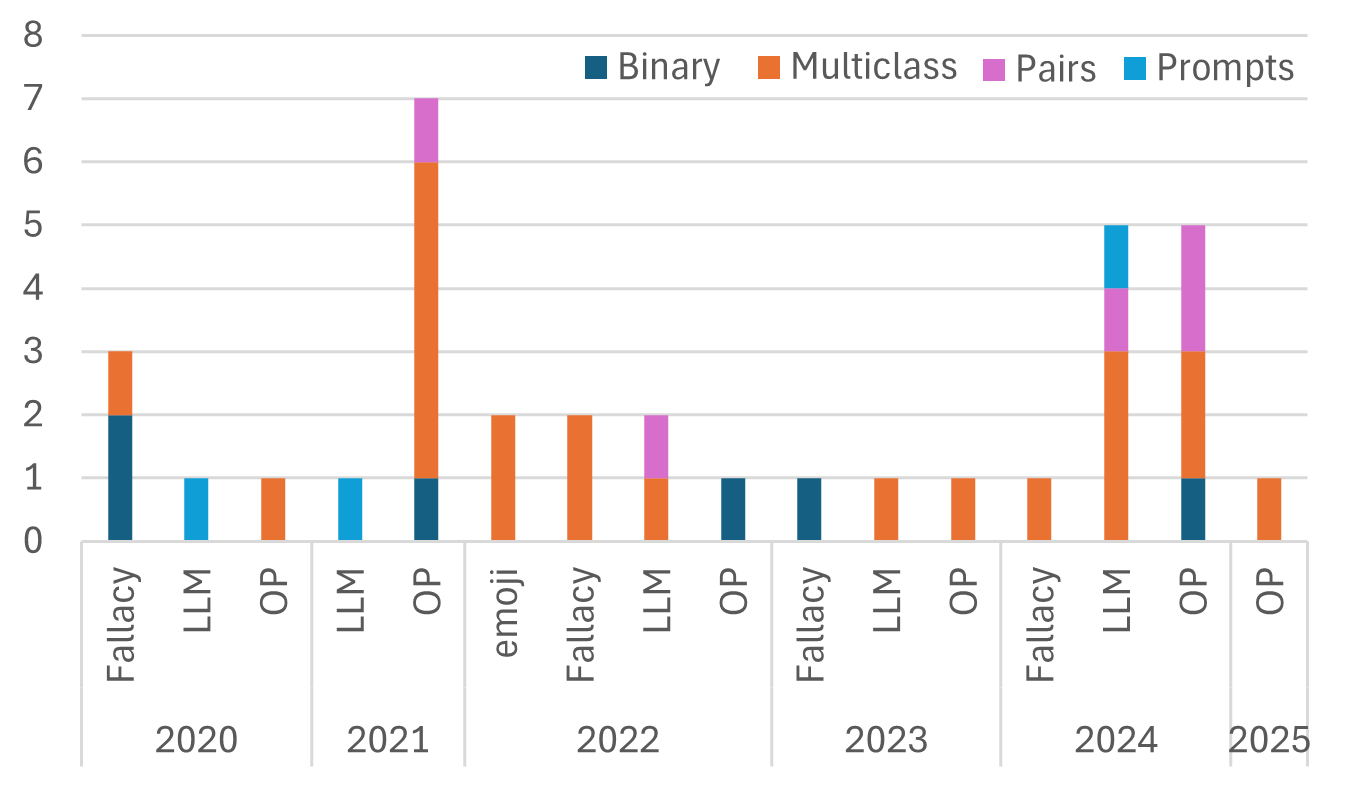}
    \caption{Summary of Toxicity Datasets creation over the last five years for Fallacy, LLMs, Online Platforms (OP), and Emoji.  The Y-axis is the number of datasets.}
    \label{fig:DatasetSummary}
\end{figure}
\begin{figure}
    \centering
    \includegraphics[width=1\linewidth]{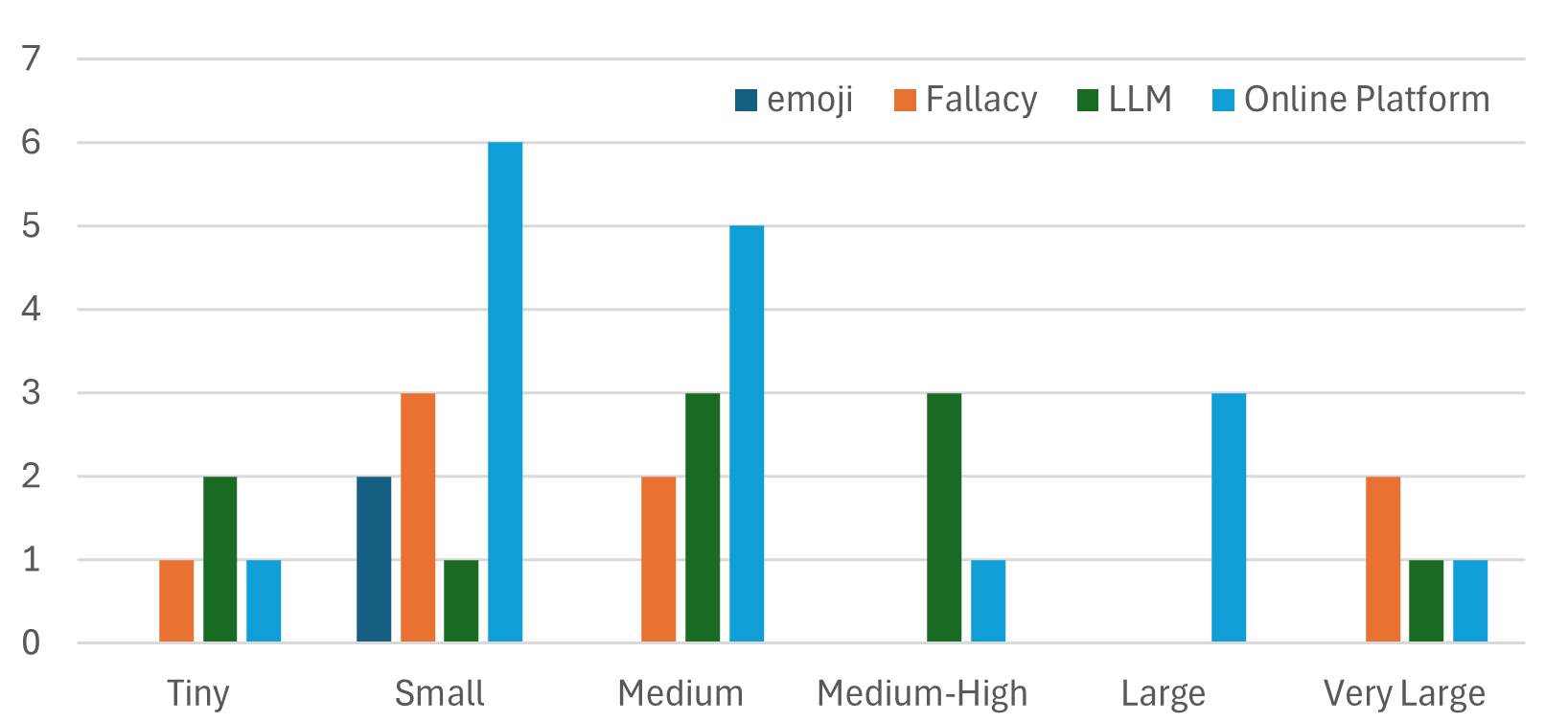}
    \caption{Datasets size distribution: Y-axis is number of datasets, and X-axis is dataset size range where Tiny is <1000, Small is between 1000 and 9999, Medium is between 10000 and 39999, Medium-High is between 40000 and 129999, Large is between 130000 to 299999, and Very Large is 300000 and above.}
    \label{fig:Datasetsize}
\end{figure}

\begin{table*}[ht!]
    \centering
    \begin{tabular}{|l|c|c|c|c|l|}
        \hline
        &&\textbf{Multiclass}& \textbf{Number} & & \\
        &&\textbf{/Binary}&\textbf{of }&&\\
        \textbf{Dataset} & \textbf{Year} & \textbf{/Both} & \textbf{classes} & \textbf{Size} & \textbf{Purpose} \\
        \hline

        HateBenchSet \cite{shen2025hatebench} & 2025 & Multiclass & 34 & 7838 & benchmarking detectors for LLM- \\
        &  &  &  &  & -generated content\\
        \hline
        
        AEGISSAFETYDATASET\cite{ghosh2024aegis}& 2024 & Multiclass & 15 & 26,000 & Adaptive content safety in LLM \\
        \hline
        
        BeaverTails \cite{ji2024beavertails}&2024& Multiclass&14&360K& QA Pairs for Safety of LLMs \\
        \hline
        
        TwitterHateSpeech \cite{verhoeven2024more} & 2024 & Multiclass & 3 & 24.8K & Explicit hate detection\\
        \hline
        
        DynaCounter \cite{chung2024effectiveness} & 2024 & - & - & 1911 & Abuse and reply pair with target group \\
        &&&&& label specific to Football Premier league\\
        \hline
        
        IntentCONANv2 \cite{hengle2024intent} & 2024 & - & - & 3488 & 4 counterspeech for each hate speech\\
        &&&&& instance of positive, informative,\\ &&&&& questioning and denouncing \\
        \hline
        
        Hatemoderate \cite{zheng2024hatemoderate} & 2024 & Binary & 2 & 7.6K & Hate speech policy or rules compliance\\
        \hline
        
        Do-Not-Answer \cite{wang2024Donotanswer} & 2024 & Multiclass & 6 & 939& Suitable for LLMs safety (referred as \\
        &&&&& toxicity in this paper) training \\
        \hline
        
        Multilevel Annotated Fallacy &2024& Both& 23 &9745& 2 classes at Level 0, 3 classes at Level 1\\
        (MAFALDA) \cite{helwe2024mafalda} &&&&& and 23 classes at level 3 \\
        \hline
        
        Thoroughly Engineered &2024&-&-&2546&Dataset of prompts to break the \\ 
        Toxicity (TET) \cite{luong2024ThoroughlyEngineeredToxicity} &&&&& protective layers of LLMs \\ 
        \hline
        
        SafeEdit \cite{wang2024detoxifyingwithSafeEdit} & 2024 & Multiclass & 9 & 8100 & Detection and detoxification in web or \\
        &&&&& SMP SafeEdit\\
        \hline
        
        MultiParaDetox \cite{dementieva2024multiparadetox} & 2024 & - & - & 16K & Extending ParaDetox, supports \\
        &&&&& Russian, Ukranian and Spanish\\
        \hline

        CivilComments \cite{duchene2023benchmark}&2023& Both&25&1.8M&Unintended bias Classification\\
        \hline
        
        PolitiFact-Oslo Corpus \cite{poldvere2023politifact-oslo} &2023& Binary&2&2745& Fake news analysis and detection \\
        \hline
        
        HarmfulQ \cite{shaikh2023secondHarmfulQ} & 2023 & Multiclass & 6 & 200 & Prompts, zero-shot chain of thoughts  \\
        &&&&& increases bias and toxicity with model size \\
        \hline

        TOXIGEN \cite{hartvigsen2022toxigen} & 2022 & Binary & 2 & 274,186 & Detection in online platforms \\
        \hline
        
        TruthfulQA \cite{lin2022truthfulqa}& 2022& Multiclass& 4& 817& 817 questions that span 38 categories\\
        \hline
        
        FACtuality \& pOlitical bIas & 2022 & Multiclass&65& 3.3M& 4,150 news-spreading users with\\
        Dataset (FACTOID) \cite{sakketou2022factoid} &&&&& 3.3M Reddit posts\\
        \hline
        
        HateMojiCheck \cite{kirk2022hatemoji} &2022&Multiclass&7& 3930& Detection of emoji based Toxicity\\
        \hline
        
        HateMojiBuild \cite{kirk2022hatemoji} & 2022&Both&54& 5,912 & Hate/non-hate; 54 target groups\\
        \hline
        
        HolisticBias \cite{smith2022mHolisticBias} &2022&Multiclass&600&45K& Prompts dataset expanded to \\ 
        &&&&& 769 descriptors \\
        \hline
        
        ParaDetox \cite{logacheva2022paradetox} &2022&-&-&10K& English toxic \& paraphrase pairs in LLMs\\
        \hline
        
        Implicit Hate Speech Corpus \cite{elsherief2021latenthatred} & 2021& Both & 7 & 22K & Implicit hate detection\\
        \hline
        
        HateXplain* \cite{mathew2021hatexplain} & 2021 & Multiclass & 3 & 20K & Implicit hate detection with 10 \\
        &&&&& target group labels \\
        \hline
        
        DynaHate\# \cite{vidgen2021learningDynahate} & 2021 & Binary & 2 & 40K & Implicit hate detection; Hate class with\\
        &&&&& 6 labels and 29 target group labels\\
        \hline
        
        Bias in Open-Ended Language & 2021 & Multiclass& 5&23,679& 5 domains with total of 43 target\\
        Generation Dataset (BOLD) \cite{dhamala2021bold} &&&&&  groups\\
        \hline
        
        Multitarget-CONAN \cite{fanton2021humanMultitargetCONAN} &2021&-&-& 5000 & Hate speech and counter Narrative pairs \\
        \hline
        
        Bot Adversarial Dialogue &2021&-&-&70K& ChatBot safety 5K dialogues and\\ 
        (BAD) \cite{xu2021botBAD} &&&&& 70K utterances\\
        \hline

        RedditBias \cite{barikeri2021redditbias}&2021& Multiclass & 4 & 11,873 & Bias on 4 dimensions: religion, race, \\
        &&&&& gender and orientation \\
        \hline

        Online Misogyny EACL2021 \cite{guest2021Misogyny} & 2021 & Multiclass & 5 & 6.567K & Online Misogyny collected from Reddit\\
        &&&&&  posts, annotated by trained annotators\\
        \hline
        
        Social Bias Inference  & 2020 & Multiclass & 7 & 150K & Implicit hate with 7 variables; up to 3\\
        Corpus* \cite{sap2020socialBias}&&&&& multi-labels; target group having 16 labels\\
        \hline
        
        \end{tabular}
    \caption{Toxicity Datasets; [* Multi-class and Multi-label for each class and target group, \# Binary with Multi-label for Hate class and Target group, @ Binary with Two Labels each, - Neither binary nor multiclass (Question-answer type dataset)]}
    \label{tab:Toxicity_Datasets}
\end{table*}

\begin{table*}[ht]
    \ContinuedFloat
    \centering
    \begin{tabular}{|l|c|c|c|c|l|}
        \hline
        &&\textbf{Multiclass}& \textbf{Number} & & \\
        &&\textbf{/Binary}&\textbf{of }&&\\
        \textbf{Dataset} & \textbf{Year} & \textbf{/Both} & \textbf{classes} & \textbf{Size} & \textbf{Purpose} \\
        \hline

        CoAID @ \cite{cui2020coaid} & 2020 & Binary & 2 & 5.2K & Misinformation regarding COVID with\\
        &&&&& fake or real classes \\
        \hline
        
        RealToxicityPrompts \cite{gehman2020realtoxicityprompts} & 2020 & - & - & 100K& Sentence paired with toxicity scores\\
        \hline
        
        PolitiFact \cite{garg2020politifact} &2020& Multiclass&6&21K& Dataset of political enws\\
        \hline
        
        FakeNewsNet \cite{shu2020fakenewsnet} &2020& Binary&2&2.67B& Repository of Politifact and GossipCop\\
        \hline 
        
        Jigsaw Toxic Comment & 2018 & Multiclass & 6& 159K & Curation of Wikipedia Comments for\\
        Classification Challenge \cite{van2018challenges} &&&&& Kaggle hosted competition in 2018 \\
        \hline
        
        GossipCop \cite{Gossipcop} & 2009 & Binary & 2 & 22.152K & Fake or Real; subset from a fact-checking\\
        &&&&&  website created in 2009\\
        \hline





    \end{tabular}
    \caption{Toxicity Datasets (-- continued from previous page)}
    \label{tab:Toxicity_Datasets}
\end{table*}
 Toxicity datasets used in surveyed research are collated in Table \ref{tab:Toxicity_Datasets} and is summarized in Figure \ref{fig:DatasetSummary}.
The datasets used vary by application domains and research objectives. The datasets are usually curated for specific application domains. Thus, some are binary datasets \cite{hartvigsen2022toxigen, vidgen2021learningDynahate, Gossipcop, cui2020coaid, zheng2024hatemoderate, poldvere2023politifact-oslo} whereas others are multiclass. Many datasets are multi-label where the examples have primary hate classifications like \textit{hate} and \textit{not hate}, secondary labels for the target class, and additional fine-grained labels \cite{ghosh2024aegis,ji2024beavertails, elsherief2021latenthatred, sap2020socialBias, mathew2021hatexplain,verhoeven2024more,van2018challenges,wang2024Donotanswer,lin2022truthfulqa,dhamala2021bold, sakketou2022factoid,garg2020politifact, kirk2022hatemoji, wang2024detoxifyingwithSafeEdit,smith2022mHolisticBias, shen2025hatebench, shaikh2023secondHarmfulQ, barikeri2021redditbias, helwe2024mafalda}. AEGISSAFETYDATASET \cite{ghosh2024aegis},  BeaverTails \cite{ji2024beavertails}, Do-Not-Answer \cite{wang2024Donotanswer}, HolisticBias \cite{smith2022mHolisticBias}, and HarmfulQ \cite{shaikh2023secondHarmfulQ} are datasets used in toxicity detection and detoxification in LLMs. These are multiclass datasets curated for LLM safety. Thoroughly Engineered Toxicity (TET) \cite{luong2024ThoroughlyEngineeredToxicity}, MultiParaDetox \cite{dementieva2024multiparadetox}, ParaDetox \cite{logacheva2022paradetox}, RealToxicityPrompts \cite{gehman2020realtoxicityprompts}, and RedditBias \cite{barikeri2021redditbias} are toxic and paraphrase pairs datasets for LLMs safety. TRam1/safe-guard-prompt-injection\footnote[1]{https://huggingface.co/datasets/xTRam1/safe-guard-prompt-injection}, deepset/prompt-injections\footnote[2]{https://huggingface.co/datasets/deepset/prompt-injections} are Prompt Injection datasets, and lmsys/toxic-chat\footnote[3]{https://huggingface.co/datasets/lmsys/toxic-chat}, SetFit/toxic\_conversations\_50k\footnote[4]{https://huggingface.co/datasets/SetFit/toxic\_conversations\_50k} are toxicity datasets from Huggingface. TruthfulQA 
\cite{lin2022truthfulqa} is a dataset used to check truthful answers generated by LLM for the questions. BAD \cite{xu2021botBAD} is a dataset for chatbot safety. 
TOXIGEN \cite{hartvigsen2022toxigen}, Implicit Hate Speech Corpus \cite{elsherief2021latenthatred}, Social Bias Inference Corpus \cite{sap2020socialBias}, HateXplain \cite{mathew2021hatexplain}, DynaHate \cite{vidgen2021learningDynahate}, TwitterHateSpeech \cite{verhoeven2024more}, HateMojiCheck \cite{kirk2022hatemoji}, HateMojiBuild \cite{kirk2022hatemoji}, CivilComments \cite{duchene2023benchmark}, and Jigsaw Toxic Comment Classification Challenge \cite{van2018challenges} are the datasets for toxicity detection in the web/SMP context. 
DynaCounter \cite{chung2024effectiveness}, IntentCONANv2 \cite{hengle2024intent}, Hatemoderate \cite{zheng2024hatemoderate}, BOLD \cite{dhamala2021bold}, SafeEdit \cite{wang2024detoxifyingwithSafeEdit}, Multitarget-CONAN \cite{fanton2021humanMultitargetCONAN}, and HateBenchSet \cite{shen2025hatebench} are datasets for toxicity detection and detoxification in web/SMP. 
GossipCop\footnote[5]{https://huggingface.co/datasets/osusume/Gossipcop}, CoAID \cite{cui2020coaid}, FACTOID \cite{sakketou2022factoid} MAFALDA \cite{helwe2024mafalda}, PolitiFact \cite{garg2020politifact} PolitiFact-Oslo Corpus \cite{poldvere2023politifact-oslo}, FakeNewsNet \cite{shu2020fakenewsnet} are datasets for fallacy detection.
As shown in Figure \ref{fig:Datasetsize}, most of the datasets are small and cannot be used in a real-world application. Only 11 datasets namely, RealToxicityPrompts, Bot Adversarial Dialogue, HolisticBias, DynaHate, Jigsaw Toxic Comment Classification Challenge, Social Bias Inference Corpus, TOXIGEN, FakeNewsNet, FACtuality \& pOlitical bIas, BeaverTails, and CivilComments, can be used for future research, providing generalization. Out of these, only two of them, \textit{RealToxicityPrompt} include implicit toxicity and \textit{Beavertails} for explicit toxicity, can be used for detoxification. 
The remaining smaller datasets are mainly curated to prove the toxicity problems and call for further research.

\begin{figure}[h]
\centering
\includegraphics[width=\linewidth]{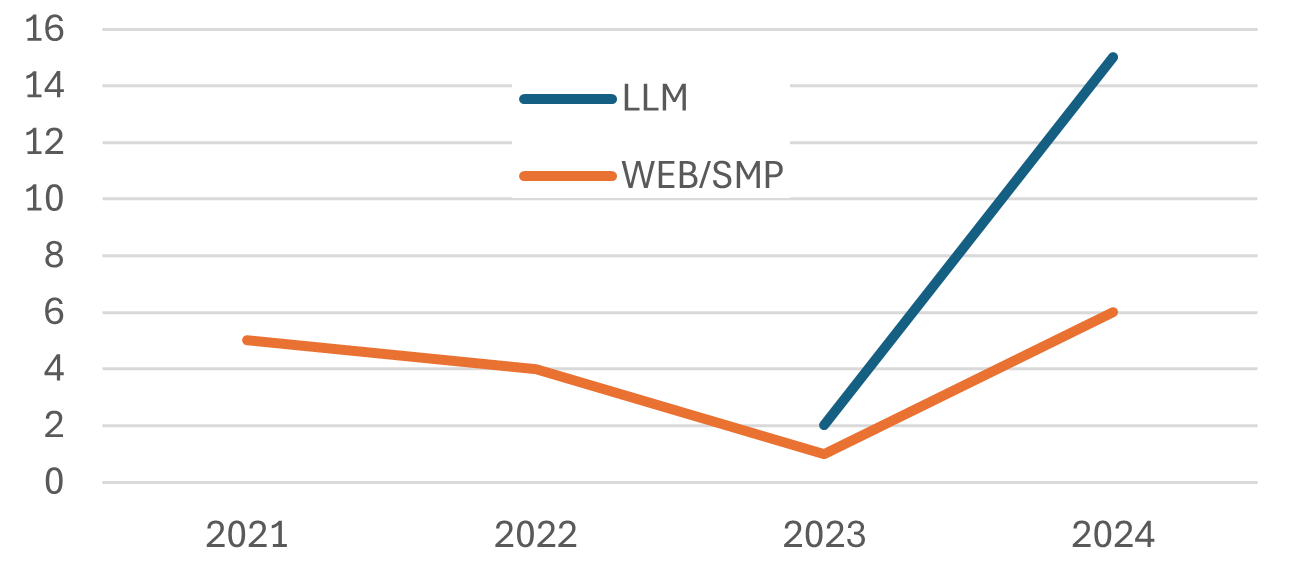}
\caption {The involvement of LLMs in Toxicity Detection and Mitigation has been increasing rapidly since 2023, compared with other methods.}
\label{fig:Trend_LLM}
\end{figure}

\begin{table*}[ht!]
    \centering
    \begin{tabular}{|l|c|c|c|c|c|c|c|c|c|l|}
        \hline
         \textbf{Toxicity Research} & \textbf{Year} & \textbf{NDS} & \textbf{NFW} & \textbf{NMT} & \textbf{UEM}& \textbf{Domain} & \textbf{Mode} & \textbf{DET} & \textbf{MTG} & \textbf{Remarks}\\
         \hline
         HateBench \cite{shen2025hatebench} & 2025 & Yes & Yes & No & Yes & WEB & Text & Yes & No & Explicit\\
         \hline

         Ismail et al. \cite{ismail2025enhancing} & 2025 & No & Yes & Yes & No & GP & Text & Yes & No & Implicit\\
         \hline
         
         Zheng et al. \cite{zheng2025lightweight} & 2025 & No & No & Yes & Yes & LLM & Text & Yes & Yes & Explicit\\ 
         \hline

         ToxEdit \cite{zhang2025toXedit} & 2025 & No & No & Yes & Yes & LLM & Text & Yes & Yes & Explicit \\
         \hline
         
         LATTE \cite{koh-etal-2024-llms}& 2024 & No & Yes & No & Yes & LLM & Text & Yes & No & Both \\
         \hline
         
         SEAS \cite{diao2024seas}& 2024 & Yes & Yes & No & Yes & LLM & Text & Yes & Yes & Explicit \\
         \hline
         
         Goodtriever \cite{ermis2024one} & 2024 & No & Yes & No & Yes & Any & Text & Yes & Yes & Explicit\\ 
         \hline

         Fu et al. \cite{fu2024cross} & 2024 & Yes & No & No & Yes & LLM & Text & Yes & Yes & Explicit \\
         \hline

         ADPO \cite{kim2024adversarial} & 2024 & No & No & Yes & Yes & LLM & Text & Yes & Yes & Explicit\\
         \hline

         CF-Detox$_{tigtec}$ \cite{bhan2024mitigating} & 2024 & No & No & Yes & No & Any & Text & Yes & Yes & Explicit\\
         \hline

         Lin et al. \cite{lin2023towards} & 2024 & No & Yes & No & Yes & LLM & Text & Yes & Yes & Explicit\\
         \hline

         RIDERS \cite{li2024focus} & 2024 & No & No & Yes & Yes & LLM & Text & Yes & Yes & Explicit\\ 
         \hline

         DETOXIGEN \cite{niu2024parameter} & 2024 & No & Yes & Yes & No & LLM & Text & Yes & Yes & Explicit \\
         \hline
         
         CoARL \cite{hengle2024intent} & 2024 & No & Yes & No & Yes & Web & Text & Yes & Yes & Explicit \\ 
         \hline

         MICo \cite{siegelmann2024mico} & 2024 & Yes & No & Yes & Yes & LLM & Text & Yes & Yes & Explicit \\
         \hline

         Schafer et al. \cite{schafer2024hierarchical} & 2024 & No & No & Yes & No & Web & Text & Yes & Yes & Explicit \\
         \hline

         Wang et al. \cite{wang2024detoxifying} & 2024 & Yes & No & Yes & Yes & LLM & Text & Yes & Yes & Explicit \\
         \hline
         
         ToxCL \cite{hoang2024toxcl}& 2024 & No & Yes & Yes & No & SMP & Text & Yes & No & Implicit \\ 
         \hline

         Verhoeven et al. \cite{verhoeven2024more}& 2024 & No & No & Yes & No & SMP & Text & Yes & No & Explicit\\
         \hline
         

         BeaverTails \cite{ji2024beavertails}& 2024 & Yes & No & No & Yes & LLM & Text & Yes & Yes & Explicit \\
         \hline

         SELF-GUARD \cite{wang2024selfguard}& 2024 & No & No & Yes & Yes & LLM & Text & Yes & Yes & Explicit \\
         \hline
         
         AEGIS \cite{ghosh2024aegis}& 2024 & Yes & Yes & Yes & No & LLM & Text & Yes & Yes & Both \\
         \hline

         Chung et al. \cite{chung2024effectiveness}& 2024 & Yes & No & Yes & No & SMP & Text & Yes & Yes & Explicit\\
         \hline

         GuardFormer \cite{o2024guardformer} & 2024 & Yes & No & Yes & Yes & LLM & Text & Yes & Yes & Both\\
         \hline

         SharedCon \cite{ahn2024sharedcon} & 2024 & No & No & Yes & No & WEB/SMP & Text & Yes & No & Implicit\\
         \hline

         Hu et al. \cite{hu2024bad} & 2024 & No & No & Yes & No & LLM & Text & Yes & Yes & FN\&M\\
         \hline
         
         BiasAlert \cite{fan2024biasalert} & 2024 & No & No & No & Yes & LLM & Text & Yes & No & Both\\ 
         \hline
         
         DisCGen \cite{hassan2023discgen} & 2023 & Yes & Yes & No & Yes & LLM & Text & Yes & Yes & Both \\
         \hline
         
         Markov et al. \cite{markov2023Moderation}& 2023 & No & Yes & Yes & No & WEB & Text & Yes &  No & Explicit\\
         \hline
         
         NeMo \cite{rebedea2023nemo}& 2023 & No & Yes & Yes & No & LLM & Text & Yes & Yes & Both \\
         \hline
         
         HateMoji \cite{kirk2022hatemoji}& 2022 & Yes & No & No & Yes & SMP & Text & Yes & No & Explicit\\
         \hline
         
         Factoid \cite{sakketou2022factoid} & 2022 & Yes & No & No & Yes & WEB/SMP & Text & Yes & No & FN\&M* \\
         \hline
         
         HSBERT \cite{toraman2022HSBERT} & 2022 & Yes & No & No & Yes & SMP & Text & Yes & No & Explicit \\
         \hline

         CapsNet– & 2022 & Yes & No & Yes & No & SMP & Text & Yes & No & Explicit \\
         ConvNet \cite{kumar2022multimodalcyberbullying} & & & & & & & Image/InfoG & & & \\
         \hline
         
         HateXplain \cite{mathew2021hatexplain}& 2021 & Yes & No & Yes & Yes & SMP & Text & Yes & No & Explicit\\
         \hline

         LFTW \cite{vidgen2021LFTW}& 2021 & No & Yes & Yes & No & SMP & Text & Yes & No & Both\\
         \hline

         Guest et al. \cite{guest2021Misogyny} & 2021 & Yes & No & No& Yes & SMP & Text & Yes & No & Explicit   \\
         \hline

         Raj et al. \cite{raj2021cyberbullying}& 2021 & No & Yes & No & No & Web & Text & Yes & No & Explicit\\
         \hline
         
         RedditBias \cite{barikeri2021redditbias}& 2021 & Yes & Yes & No & Yes & Web/SMP & Text & Yes & Yes & Both\\
         \hline


    \end{tabular}
    \caption{Summary of toxicity detection and mitigation research presented in this survey. NDS = Novel Dataset; NFW = Novel Framework; NMT = Novel Model or Technique; UEM = Used Established Models/Benchmarks; WEB = Content hosting websites like Wikipedia; InfoG = Info-Graphics; GP = Gaming Platforms; DET = Detection; MTG = Mitigation; FN\&M* = Fake News and Misinformation }
    \label{tab:detection}
\end{table*}
\subsection{Toxicity Detection}
Toxicity detection is a process of implementing a mechanism to identify a toxicity incident in communication in a given environment. Detection techniques may or may not be LLM-based, depending on the platforms, although the trend seems to move towards LLM-based approaches as shown in Figure \ref{fig:Trend_LLM}. In this paper, both LLM-based and non-LLM-based detection techniques are reviewed and presented in Table \ref{tab:detection}. It is evident that different LLMs have different definitions of toxicity and approaches for detection \cite{shen2025hatebench}. Koh et al. \cite{koh-etal-2024-llms} proposed a framework for evaluating toxicity with variable definitions, emphasizing the flexibility of the toxicity metric and demonstrating the adaptability of LLMs to prompts. To train toxicity detectors, some researchers have generated datasets like TOXIGEN, AEGISSAFETYDATASET, HateBenchSet, and HateXplain (refer to Toxicity Dataset Table \ref{tab:Toxicity_Datasets}). Recently, many researchers have generated LLM-based techniques and frameworks. A notable approach called ToxCL \cite{hoang2024toxcl}, brings a new perspective by not just detecting toxicity, but also explaining the detected instance. Explainable AI (XAI) is a research area to build transparency and trustworthiness of a model \cite{mersha2025evaluating, mersha2025unified, yigezu2024ethio}. It increases detection performance compared to developing separate models for detection and explanation. 
In the following sections, we have listed the broader types of toxicity research. 

\subsubsection{Detection of Explicit Toxicity}
A recent study conducted by Shen et al. \cite{shen2025hatebench} on the detection of explicit toxicity in LLM-generated web content highlights limitations in the detection techniques that enable adversarial attacks on LLMs with greater efficiency. HateBench \cite{shen2025hatebench} is a novel framework to create a dataset dynamically. It utilizes an ensemble of 8 models, out of which 4 are research (LFTW, TweetHate \cite{antypas2023TweetHate}, HSBERT, and Bert-HateXplain) and 4 are commercial models (Perspective, Moderation, Detoxify-original, and Detoxify-unbiased). 
The study curated a dataset and proposed a detection mechanism for web content. The purpose of this research is to highlight that it is easy to launch hate campaigns without being detected, calling for a rigorous research effort in LLM Safety and detoxification. Although this framework cannot be used in real-world toxicity mitigation, it provides a base for future research. 

Verhoeven et al. \cite{verhoeven2024more} proposed another toxicity detection mechanism in SMP using Community models' generalization via graph meta-learning. They have trained the model on GossioCop and evaluated on CoAID and TwitterHateSpeech. They have pointed out that such models are either transductive (assumption: graphs are static and do not change) or inductive (assumption: graphs change all the time). They have pointed out the absence of datasets in inductive social graphs and achieved generalization using graph meta-learning. 
Although their results are encouraging, they have pointed out that further work is required to allow such a solution to be implemented in real-world applications. 

Markov et al. \cite{markov2023Moderation} proposed a detection mechanism (Moderation API) for web content on a 219K dataset of real-world, synthetic, and production content. They have acknowledged that their model suffers from bias, fairness, robustness, and multilingual support issues and requires further red-teaming and active learning experiments. Their dataset annotation is based on their taxonomy of eight categories and further subcategories of undesired content or toxicity. Imbalance of data categories is another problem to solve; using data augmentation is suggested. 

In developing HateMoji \cite{kirk2022hatemoji} Kirk et al. curated two datasets, HateMojiCheck and HateMojiBuild, and proposed a detection mechanism for SMP where emojis are used in hateful language expressions. While they have curated a multiclass dataset, their experiments for hate detection detect binary level only. Their experiments only included six identities and had limited scope. 

In creating HSBERT \cite{toraman2022HSBERT} Toraman et al. curated a dataset of 100K instances for English and Turkish, each with scalability, and a proposed detection mechanism on SMP. These datasets cover only 5 subcategories of gender, race, religion, politics, and sports. Thus, limited in scope. In CapsNet–ConvNet \cite{kumar2022multimodalcyberbullying}, Kumar et al. curated a dataset of 10K instances from Twitter, YouTube, and Instagram and proposed a detection mechanism for cyberbullying on SMP.  They have proposed a deep learning based model for toxicity detection for multi-modal data (text, image, and image with text). Due to the dynamic landscape of SMP, and Natural Language Processing (NLP) models' capability to outperform deep learning models, further research in cyberbullying detection is suggested using state-of-the-art methods

While developing HateXplain \cite{mathew2021hatexplain}, Mathew et al. curated a multi-label dataset and proposed a detection mechanism using Ground Truth Attention with Filter for ten target groups on SMP. HateXplain dataset has three labels: one for Hate Speech, Offensive, Normal; the second for the target group; and the third for the rationale. This work shows that the models that perform great at first or second label classification cannot perform as well at third rationale explanation tasks. 

Guest et al. \cite{guest2021Misogyny} curated a 6,567-post dataset for online misogyny, misogyny taxonomy, and proposed a binary detection mechanism for SMP. Their dataset is generated from 11 weeks of Reddit posts, annotated by trained annotators, and is publicly available. Their dataset is highly imbalanced in binary and across fine-grained misogyny, which prevents them from classifying the fine-grained using their mechanism, given the small size of their dataset. But their taxonomy and high-quality annotation pave the path for future work in detecting misogyny. 

Raj et al. \cite{raj2021cyberbullying} propose an architecture for detecting cyberbullying in online platforms. They have experimented with various machine learning algorithms and neural network models on two Wikipedia datasets. They found that a neural network bidirectional Gated Recurrent Unit with Global Vectors outperforms existing state-of-the-art techniques. This technique can be experimented on other datasets to compare their performance with the current state-of-the-art. Another observation of this work, which might be questionable, is their datasets' imbalance, having less than 9-12\% of toxic classifications. Neural networks' classification will tend to be towards the non-toxic side. If the performance parameters are shown for only the cyberbullying and toxic class would make their conclusion more reliable.

\subsubsection{Detection of Implicit Toxicity}

Ismail et al. \cite{ismail2025enhancing} have highlighted that the severe consequences of aggressive toxicity in online gaming platforms include suicide, and sensed the urgency for their detection. They have proposed ``a novel embeddings-based valence lexicon approach'' using natural language processing, to detect it on a Twitch Corpus, specifically for gaming platforms. They have also suggested that their approach should work on other online platforms as well. While this approach detects the implicit toxicity, its further target classification or explanation is not generated.  

In ToxCL \cite{hoang2024toxcl}, Hoang et al. proposed a framework of detection mechanism for an SMP with a Target Group Generator, an encoder-decoder for a detection teacher classifier, and a knowledge distillation approach for explanation. Thereby identifying the toxicity class, target group, and explaining their output. They have utilized target group labels in \textit{Implicit Hate Corpus} and \textit{SBIC
} datasets, and nineteen fine-grained categories from HateXplain dataset. They have performed an effectiveness evaluation by comparing ToxCL with human evaluation. They have noted that their framework may not perform well on abbreviated texts, symbols, sarcasm, and irony; and may have more than one non-overlapping explanation.

In SharedCon \cite{ahn2024sharedcon}, Ahn et al. proposed a detection mechanism using shared semantics by clustering and contrastive learning for SMP and web content. They have used K-Means clustering and chosen a post randomly, which is close to the centroid as the shared semantic sentence in the cluster. While this method eliminates the human in evaluating the implicit toxic nature, the basis of the choice of clustering and the number of clusters is not discussed. Selecting the most suitable clustering algorithm for a specific dataset is crucial, as the effectiveness of this approach heavily depends on identifying optimal clusters, which serve as the core of this mechanism.

\subsubsection{Detection of Both Implicit and Explicit Toxicity} \label{det_Impl_Expl}
Learning from the worst (LFTW) \cite{vidgen2021LFTW} proposed by Vidgen et al. is a dynamic dataset creation mechanism for robust toxicity detection model training for SMP. It produces a balanced dataset across hate and non-hate categories. Their mechanism introduces perturbations after each round, for further challenging set creation, requiring trained annotators. They have pointed out that this technique requires specific infrastructure, cross-disciplinary domain expertise, and is time-consuming process. 

Ghosh et al. \cite{ghosh2024aegis} presented a taxonomy with 13 critical and 9 sparse content safety risk categories in the human-LLMs conversation, curated AEGISSAFETYDATASET, and developed an online adaptive content moderation technique called AEGISSAFETYEXPERTS for LLM. During the publication, they acknowledged that their dataset was not complete, and further training of AEGIS is required. With a dataset size 26K, their performance outperformed existing state-of-the-art. 

O' Neill et al. \cite{o2024guardformer} proposed a synthetic data pipeline generation approach with multi-task learning and a guardrail classifier called GuardFormer for LLM. They have a synthetic data generation LLM for given prompts, which generates corresponding safe and unsafe prompts, rationale, and labels. Followed by custom policy guardrailing by specific model \textit{PolicyGuard} and multi-policy guardrailing model \textit{GuardFormer} capable of being effective on all policies.
It offers a smaller memory footprint under 512MB and is faster, outperforming state-of-the-art on prompt injection, toxicity classification (fine-tuned), and content-safety. While their performance on 8 datasets and 11 high-performing LLMs is superior, testing it on real-world scenarios would be further helpful.

Fan et al. proposed BiasAlert \cite{fan2024biasalert}, a plug-and-play tool using Retrieval-Augmented Generation (RAG) knowledge for bias detection in LLM. They have used their metric of efficacy score, classification score, attribution score, and overall score to evaluate and compare their performance. It is evaluated on the SBIC dataset only and requires an updated RAG database corresponding to the prompt-response generation.

Koh et al. proposed LATTE \cite{koh-etal-2024-llms}, a framework that demonstrates the dynamic adaptability of LLMs to various toxicity definitions, thereby allowing for their metric flexibility. It moves away from static definitions and metrics, leaning towards the socio-cultural aspects. They have noted that it may be subject to inductive bias. As discussed in Section \ref{Sec_GenAI}, this makes LLMs vulnerable to toxicity detection issues.

\subsubsection{Detection of Bias and Fallacy}
Factoid \cite{sakketou2022factoid} is a dataset of fake news and misinformation by Sakketou et al. They proposed a detection mechanism for SMP and web content using a Graph Attention Network (GAN). This dataset's users are binary annotated for misinformation spreaders or real news spreaders, and have four fine-grained labels with their ranges of factuality degree (-3, 3), political bias (-3, 3), scientific belief (-1, 1), and satire degree (0, 1). It is one such user-centric large dataset. Their proposed models' best performances range between 58.4 to 66.2\% on F1-Scores, which have scope of improvement.

Hu et al. \cite{hu2024bad} proposed an LLM-based technique called adaptive rationale guidance (ARG) network for fake news and misinformation detection, by encoding textual description and commonsense rationales for classification. They have evaluated LLM-only, small language model (SLM) only and LLM+SLM models. And found that LLM+SLM with ARG outperforms others. While their model does well in real news detection, it struggles in fake news detection. They have highlighted that their model relies on LLM-generated rationale only; there might be other perspectives outside of this rationale that should be considered. 

For Bias detection, BiasAlert is proposed by Fan et al. is detailed in Section \ref{det_Impl_Expl}.



\subsection{Toxicity Detoxification}

Detoxification is the mechanism to remove the toxicity. 
It is performed after toxicity detection. In online platforms, for implicit or explicit toxicity with or without bias. Detoxification either uses pairs of hate speech (toxic language) and counterspeech as a dataset to generate non-hate speech non-toxic language with the same semantics or intentions, thereby replacing toxic text, detoxifying the content. Corresponding research summarization is presented in Table \ref{tab:detection}.

\subsubsection{Detoxification of Explicit Toxicity in LLM}
Zheng et al. \cite{zheng2025lightweight} proposed a lightweight fine-tuned Sentence-BERT-based guardrails model for LLM safety. It reduces the parameter size to 67M compared to LLM-based (7B parameters) solutions while maintaining their performance comparable to state-of-the-art. But it is not customizable and flexible for few-shot topic filtering. 

Another safeguarding model proposed by Zhang et al. is called ToxEdit \cite{zhang2025toXedit}, which uses toxicity-aware knowledge editing. It detects toxicity dynamically while avoiding over-editing of non-toxic responses on SafeEdit. They have performed limited experiments on SafeEdit using an SVM classifier only, Neural network classification might yield better results. ToxEdit suffers from sentence repetition in its generation.

Diao et al. proposed a self-evolving adversarial safety optimization technique for LLMs called SEAS \cite{diao2024seas}. They have generated a dataset of adversarial prompts and a pipeline for dynamic safety optimization in the SEAS framework, showing performance comparable to GPT-4. They have highlighted that the quality of their framework improves with the number of iterations; hence, initial dataset diversity may be limited. They have used Llama Guard 2 in their framework for safety classification; thus, human evaluation is advisable. 

Ermis et al. proposed a framework called Goodtriever \cite{ermis2024one} comparing RAG with fine-tuned models under evolving and static language with multilingual support using translation for text generation. It lays the foundation for multilingual toxicity mitigation.
Due to a lack of low-resource languages, cultural differences, translation variation, and information loss during translation are evident.

Fu et al. \cite{fu2024cross} curated an Instruction Tuning Safety-defense dataset with 2000 documents and proposed a fine-tuned model with Low-Rank Adaptation (LoRA). It identifies malicious documents, resulting in the refusal of processing, while processing benign documents effectively. This model is trained on a balanced dataset only.

Kim et al. proposed a training algorithm called adversarial direct performance optimization (ADPO). It assigns safe or unsafe responses using the toxic control token \cite{kim2024adversarial}. This has limitations of harmfulness and bias in LLM's annotated labels. Human annotation is also prone to bias due to the same demographic profile. Further, only 16K dataset is used for this work; using 160K dataset is expected to enhance the performance.  

Lin et al. \cite{lin2023towards} proposed a framework with four AI agents, an original chatbot, with user therapist, and critic, with reinforcement learning based LLM tuning to detect and mitigate toxicity in the behavior of chatbots. But this implementation highly relies on the quality of datasets during training to enhance generalization, such that chatbot's bias and user's privacy are as per ethical considerations. AI agent as a psychotherapist needs to ensure communication with empathy to make it effective. Quantification is necessary to be evaluated. Explainable AI and reinforcement learning techniques to be explored in complicated interactions to allow cooperation. 

RIDERS \cite{li2024focus} is another mitigation technique proposed by Cao et al. that uses a toxic chain-of-thought (CoT) with residual decoding and the serial position swap. It reduces the toxic CoT problem and improves commonsense reasoning. They have focused on multiple-choice questions with open-ended commonsense reasoning, calling for benchmark-related research. Additionally, their reasoning task does not consider mathematical questions.

DETOXIGEN \cite{niu2024parameter} is another detoxifier model proposed by Niu et al., which is an ensemble of a pre-trained language model as a generator that is parameter-efficient using contrastive decoding, and a detoxifier that produces toxic tokens for the generator. They are trained on \textit{RealToxicityPrompts} dataset. This research lacks experimentation where both LLMs are trained on separate datasets and may not cover the complete toxicity scope. Additionally, their model's human evaluation and generalization study have not been conducted. 

Hengle et al. proposed CoARL \cite{hengle2024intent}, a framework using multiple instruction tuning using LoRA and RL using LLMs for counterspeech generation. While their model outperforms state-of-the-art LLMs on their metric, it is important to note that they have used a small dataset for their experiment and cannot be considered exhaustive. They have pointed out that their RL reward approach might add bias, non-alignment with human perceptions, and multiple feedback loops are not accounted for in the model, which may affect its effectiveness over time. 

Siegelmann et al. proposed MICo \cite{siegelmann2024mico}, Models with Inhibition Control for Preventative detoxification, and curated a dataset with 2850 entries. Due to the smaller size of their dataset, it is non-representative of diverse toxic content. Thus, this model can be trained on a larger dataset and may require fine-tuning and modifications.

Schafer et al. \cite{schafer2024hierarchical} proposed a technique for the mitigation of identity term bias in toxicity detection. They have defined a bias target in three levels, improving the generalization across targets. They have conducted a single run of experiments focused on religious bias and thus have not considered their scores to be robust. 

Wang et al. \cite{wang2024detoxifying} curated a dataset and proposed a technique with knowledge editing detoxification and an intra-operative Neural Monitoring mechanism. This work sets the groundwork for future research like ToxEdit \cite{zhang2025toXedit} described earlier. 

Ji et al. proposed BeaverTails \cite{ji2024beavertails}, curated a large dataset of QA pairs with meta-labels for safety, and evaluated the model with metrics of helpfulness and harmlessness. They have highlighted that the annotators' demographic diversity is limited, and there might be overlap in certain categories. 

Wang et al. proposed an approach called SELF-GUARD \cite{wang2024selfguard} to prevent jailbreaking attacks by performing content assessment and toxicity detection on prompts and responses. They have highlighted that this approach reduces harmful content generation probability but does not eliminate it. 

\subsubsection{Detoxification of Both Implicit and Explicit Toxicity in LLMs}
DisCGen \cite{hassan2023discgen} by Hassan et al. curated a counterspeech dataset and proposed a discourse-aware framework. Their classifier is trained on Multitarget-CONAN, which is a small dataset, exhibits bias \cite{tonja2023first}. They have manually relabeled the wrongly tagged instances in their counterspeech dataset.   

Rebedea et al. proposed NeMo \cite{rebedea2023nemo}, an open-source toolkit to add programmable rails to LLM-based applications for implicit and explicit toxicity. They acknowledge that it cannot be used as a standalone safety mechanism, but can be added to already existing embedded guardrails to prevent jailbreak attacks.

\subsubsection{Detoxification Explicit Toxicity in SMP and Web Content}
Chung et al. \cite{chung2024effectiveness} curated a small dataset and proposed a detoxification technique for Twitter, where the football players were targeted during the UK Premier League championship. 
Barikaeri et al. proposed RedditBias \cite{barikeri2021redditbias} and curated a dataset. 
Bhan et al. \cite{bhan2024mitigating} proposed CF-Detox$_{tigtec}$, a counterfactual-based approach for detoxifying text using local feature importance explanations. Their method preserves semantic content while reducing toxicity, demonstrating the potential of eXplainable AI to support language detoxification with interpretability and content faithfulness. 

\subsection{Model Performance Metrics and Evaluations}
Most toxicity detection and detoxification mechanisms are evaluated using the F1-Scores, accuracy, precision and recall metrics. To ensure that the model's performance is robust and fair, a generalized framework is essential. It should accommodate various fine-grained toxicities based on toxicity context, platform, and environment. 

Nazir et al. proposed an open-source Python toolkit LangTest \cite{nazir2024langtest}, with 60 tests and claimed to help improve the performance of LLMs and NLP models in regards to toxicity recognition and mitigation. It dynamically generates the automated testing workflow based on real-world applications. It trains, evaluates, runs tests, augments the training datasets, and repeats the process. Test results before and after augmentation are compared for the ``bias, robustness, accuracy, fairness, and security'' aspects of the model. 

For bias and fairness assessment of an LLM, Bouchard et al. proposed LangFair \cite{bouchard2025langfair}, a Python package. It claims to provide an evaluation framework tailored to LLM tasks, considering their responses. It generated toxicity score for the words and stereotypes based on occurrence and classification. 

Rottger et al. proposed HateCheck \cite{rottger2021hatecheck}, consisting of 29 function tests, a test suite to evaluate toxicity detection models. They targeted the annotation process of toxicity datasets. 

Verhoeven et al. \cite{verhoeven2024more} proposed an evaluation setup for the generalization capability of community models for fallacy and hate-speech detection, claiming that social subgraphs and graph meta-learning have better generalizability compared to standard community models. \\

In most of the \textit{toxicity mitigation} studies, where the dataset is curated or synthesized, a common problem is the lack of diverse annotators' backgrounds, causing bias. In many proposed techniques, the usage of a single small dataset for training and evaluation limits their scope, generalization, and robustness. Some studies using machine learning and deep learning techniques are conducted on an imbalanced dataset, which raises concerns about the reliability of their results.
Many studies focused on providing the necessary context and demonstrating the problem's relevance to pave the way for future research. 

\section{Open Challenges and Future Directions} \label{Sec_Open} 
Despite advances in toxicity detection, several challenges remain. These include the scarcity of representative datasets, difficulties in handling emojis and informal language, and the complexity of detecting multilingual and multimodal toxic content, implicit bias, and fairness issues in LLMs. Another major open challenge is the lack of interpretability in toxicity classifiers, which limits transparency and user trust. Addressing these challenges is essential for building inclusive and accountable online platforms and AI systems.


\subsection{Datasets} 
Various datasets are being used in research in different domains to train machine learning models to detect toxicity. However, the dataset sizes are not large enough to generalize it to a large set of domains. There is an opportunity to merge many of the datasets and their classes into one large central dataset, so that research with high coverage and effectiveness can be performed. Such a dataset may serve as the starting point for various lines of research studies. Antypas et al. \cite{antypas2023TweetHate} showed better performance in toxicity detection when 13 datasets of tweets were unified. During dataset unification, challenges like different formatting, class granularity such as binary or multiclass, imbalance among classes and languages were encountered. Toraman et al. \cite{toraman2022HSBERT} emphasized that the model detection performance highly depends on the quality of the dataset on which it is being trained. XAI techniques can support the construction and quality control of datasets. The labeling inconsistencies or annotation errors can be detected by analyzing model explanations. Further use of XAI-based highlighting to make more informed decisions, improving inter-annotator agreement. This is especially important given the documented biases and inconsistencies in commonly used toxicity datasets \cite{wiegand2019detection}.

\subsection{Multi-Lingual Multi-modal Toxicity}
This survey is limited to English content in textual mode. This taxonomy can be adapted to languages other than English, this remains unexplored in this survey paper. Research involving toxicity in most languages is sparse, and any toxicity taxonomy created for them could be expanded further. This survey has presented a thoughtful and detailed Toxicity Taxonomy, which should apply to them, generally speaking. 

Toxicity detection and mitigation for modalities except text, such as audio, image, video, and multi-modal toxicity, are not surveyed in this work. There is scope to survey these modal toxicities based on the given taxonomy. 

\subsection{Implicit Bias}
Recently, research in implicit bias has picked up momentum. This survey is limited to unearthing the toxicity extant in platforms. Implicit bias exhibited during the dataset annotation process also besets datasets with the bias inherent to experts or human annotators. Further, social and algorithmic bias are also not explored in this paper. Gallegos et al. \cite{gallegos2024bias} presented a survey on LLM's bias and fairness and provided future directions for mitigation.
Unintended bias in toxicity detection is challenging to address. Lopez et al. \cite{lopez2025contextsocialmedia} concluded that dependence on the annotators of the toxicity datasets makes it difficult to achieve reliability.
Research has demonstrated bias in the toxicity detection process itself \cite{gencoglu2020cyberbullying} \cite{vaidya2020empirical}, a topic not part of this survey. 

\subsection{LLMs } 
Natural language processing with emojis, reinforcement learning with human feedback, and the vulnerabilities of LLMs pose a tremendous challenge in dealing with toxicity. The different individuals perceive emojis differently \cite{zhukova2024EmojiToxicity}. Gupta et al. \cite{gupta2023emoji} highlighted that the polarity of a statement's sentiment can be inverted when a single or multiple emojis are included in the sentiment analysis. 
The use of reinforcement learning with human feedback poses a serious challenge to retaining the models' initial training. New prompts continually enable LLMs to learn and update themselves, thereby improving their performance. In the case of prompt injection and jailbreaking attacks, this may slowly coax the model toward toxicity. This is a research gap suggesting LLMs should be equipped with a mechanism to identify which prompts to learn from or which ones to ignore. 
Vulnerabilities of LLMs beyond toxicity are not explored in this survey. However, we consider prompt injection and jailbreaking as part of toxicity in LLMs. 

\subsection{Algorithmic, Social, and Evaluation Bias}
While we have showcased research on implicit and explicit toxicity and bias based on target groups, we have not focused on identifying whether toxicity has arisen due to algorithmic bias or social bias. It is necessary to further distinguish which type of algorithmic bias it is afflicted with, whether it is consciously built-in or has crept in subconsciously. Explicit or implicit, algorithmic bias is challenging to evaluate. Evaluation metrics used in research are generic and not self-reflective, suggestive of bias in itself. How to come up with an evaluation metric for toxicity, free of bias, is another huge challenge. 

\subsection{eXplainable AI}

Studies have shown that models trained on biased datasets can incorrectly classify neutral identity-related terms such as 'black' or 'gay' as toxic \cite{bender2021dangers}. XAI techniques can uncover such patterns by highlighting consistently high attribution scores for these identity terms. This insight allows us to identify and correct biased behavior in models. 
XAI improves the transparency, fairness, and effectiveness of the toxicity classification models and detoxification systems by generating explanations on their decisions \cite{mersha2024explainable}. XAI techniques, such as LIME \cite{ribeiro2016should}, SHAP \cite{lundberg2017unified}, and Integrated Gradients \cite{sundararajan2017axiomatic}, aid in identifying words that contribute to the classification of toxicity, thereby enhancing interpretability and user trust, 
providing higher accuracy and fairness in AI systems, by highlighting toxic elements in text, improving evaluation metrics, and quality of the dataset \cite{mersha2025explainable, mersha2024explainability}. XAI also enables content moderation and online safety \cite{chamola2023review, mersha2025evaluating}. 

\section{Conclusion} \label{Sec_conclusion}
The toxicity in online platforms and AI Systems continues to remain a challenge, even though there is copious research with various perspectives and contexts being conducted, due to various reasons. One, the effort is usually piecemeal and thus not holistic, and needs consolidation. Two, the technology is evolving, and so are the facets of toxicity. Adaptability to evolving technology is necessary.
Three, the platforms deal with toxicity in a reactive manner and not a proactive manner. Proactive techniques rather than reactive ones should be encouraged. Four, a framework of governance policies is not usually involved in the detection and mitigation of toxicity in digital communication. Five, holistic approaches to develop toxicity evaluation metrics have not been forthcoming. Six, as more and more AI systems are integrated into our lives, more psychological assessments must be considered for effective quality assurance. 

Addressing AI-driven toxicities requires improved content moderation strategies, enhanced model transparency, and bias-mitigation techniques, such as adversarial training, XAI, and reinforcement learning with human feedback \cite{liu2023transforming}. Additionally, as discussed in the Section \ref{sec_stakeholders}, a collaborative effort is highly warranted. Adversaries are always several steps ahead in attacking the ecosystem with toxic behavior. All stakeholders need to continuously improve mechanisms for detection and detoxification with evolving technology and omnipresent cyberattacks to mitigate the psychological impacts society is direly facing. 



\bibliographystyle{elsarticle-num}
\bibliography{bib.bib}

\end{document}